\documentclass[a4paper,fleqn]{cas-dc}

\usepackage[authoryear,longnamesfirst]{natbib}

\usepackage[authoryear,longnamesfirst]{natbib}

\usepackage{graphicx}
\usepackage{subfigure}
\usepackage{caption}  
\usepackage{adjustbox}      
\usepackage{bm}
\usepackage{placeins}  
\usepackage{algorithm,algorithmic}
\usepackage{overpic}
\usepackage{longtable}
\usepackage{booktabs} 
\usepackage{tabularx} 
\usepackage{array}  

\def\tsc#1{\csdef{#1}{\textsc{\lowercase{#1}}\xspace}}
\tsc{WGM}
\tsc{QE}

\ExplSyntaxOn
    \cs_gset:Npn \__first_footerline:
  { \group_begin: \small \sffamily \__short_authors: \group_end: }
\ExplSyntaxOff

\begin{document}
\let\WriteBookmarks\relax
\def\floatpagepagefraction{1}
\def\textpagefraction{.001}

\shorttitle{}    

\shortauthors{C. Gao et al.}  

\title[mode = title]{Topology Understanding of B-Spline Surface/Surface Intersection  with Mapper}   

\author[1]{Chenming Gao}
\author[2]{Hongwei Lin}
\cormark[1]
\author[3]{Gengchen Li}

\address{School of Mathematical Science, Zhejiang University, Hangzhou, 310027, China} 
\address{State Key Laboratory of CAD \& CG, Zhejiang University, Hangzhou, 310058, China} 

\cortext[1]{E-mail address: hwlin@zju.edu.cn (H. Lin).}

\begin{abstract}
In the realm of computer-aided design (CAD) software, the intersection of B-spline surfaces stands as a fundamental operation. 
Despite the extensive history of surface intersection algorithms, the challenge of handling complex intersection topologies persists. 
While subdivision algorithms have demonstrated strong robustness in computing surface/surface intersection and are capable of addressing singular cases,  determining the topology of the intersection obtained through these methods is a key factor for calculating correct intersection, and remains a difficult issue.
To address this challenge, we propose a Mapper-based method for determining the topology of the intersection between two B-spline surfaces. 
Our algorithm is designed to efficiently handle various common and complex intersection topologies. 
Experimental results verify the robustness and topological correctness of this method.
\end{abstract}

\begin{keywords}
Surface/surface intersection\sep
Topology structure\sep
Topology understanding\sep
Mapper\sep
Topological data analysis\sep 
\end{keywords}

\maketitle

\section{Introduction}\label{sec:Introduction}

B-spline surface intersection is one of the fundamental operations in computer-aided design (CAD) software.\   Ensuring that surface intersection operations are robust,\   efficient,\   and topologically correct is of paramount importance.\   Despite the extensive history of surface intersection algorithms,\   complex intersection topology remains a persistent challenge.\   

Subdivision method is a robust method for calculating
	 surface/surface intersections by recursively subdividing the parameter domains fo the two surfaces,
	 leading to strip-shaped point cloud in each domain \citep{lasser1986intersection}, \citep{de1996surface}, \citep{lin2013affine}.
The key factor for computing the correct intersections is to 
	understand the topological structure of the strips of point cloud correctly,
	i.e., determining the connected components, 
	the openness or closeness of each connected component, 
	and the bifurcation points.
Because the surface/surface intersection can be very
	 complicated in some cases,
	 the topology structure of the strips of point cloud is also complicated,
	 and hard to understand.

To address this challenge,\   we propose a new method based on Mapper \citep{Singh2007mapper} for understanding the topology of intersection curves between two B-spline surfaces.\    Mapper is a significant tool in topological data analysis that extracts topological features from high-dimensional data at various scales and projects them into 2D or 3D space for visualization.\   
First,\    we employ a two-step Mapper algorithm to construct the Mapper graph of the intersection points to understand its topological structure.\    Next,\    we identify two types of characteristic vertices within the Mapper graph:\   singular vertices and boundary vertices.\   
Finally,\    by removing these characteristic vertices from the Mapper graph,\    we can partition the intersection region into several simple open or closed curves.\    
In summary,\   the main contributions of this study are as follows:

\begin{itemize}
        \item Topology structure of Surface/Surface intersection is understood by Mapper at the first time,\   showing its robustness and efficiency;\
        \item A two-step mapper algorithm is developed to extract the structure of Surface/Surface intersection point sets.\  
\end{itemize}

The rest of the paper is organized as follows.\    In Section 2,\   we review related work on  surface intersection  algorithms and Mapper algorithm.\    In Section 3,\  preliminaries on the Mapper algorithm and surface intersection are introduced.\    In Section 4,\   we propose an algorithm for understanding the topology of the intersection of two B-spline surfaces.\    In Section 5,\   we verify the effectiveness of the algorithm experimentally.\    Finally,\   we conclude this study in Section 6.\

\section{Related work}\label{sec:Related work}

\subsection{Surface/Surface intersection and determination of intersection topology}

In this section,\   we briefly review related work on Surface/Surface intersection.\   
Surface intersection is a critical issue in geometric design and has been the subject of extensive research.\   In general,\   intersection methods can be categorized as follows.\   

\textit{Algebraic methods} typically involve converting one surface into its implicit equation.\    The parametric form of the other surface is then substituted into this implicit equation to compute the Surface/Surface intersection using elimination theory \citep{sarraga1983algebraic},\   \citep{manocha1991new},\   \citep{manocha1997algebraic}.\    These methods determine the topology of the intersection by identifying "characteristic points" such as border points,\   turning points,\   and singular points \citep{farouki1986characterization},\   \citep{hass2007guaranteed}.\    However,\   determining how these points are connected remains a challenge,\   as incorrect connections can lead to topological errors.\   

\textit{Tracing methods} generate intersections by stepping from initial points on a curve branch,\   following the local differential geometry of the curve \citep{bajaj1988tracing}.\    These methods require careful selection of valid initial points to ensure no branches are missed.\    Additionally,\   determining the tracking direction at singular points of the intersecting curves is challenging.\   

\textit{Lattice methods} decompose the Surface/Surface intersection problem by computing the intersection of multiple isoparametric lines from one surface to another \citep{rossignac1987piecewise}.\    However,\   the choice of mesh resolution is crucial,\   as inappropriate resolutions may result in the omission of important characteristic points.\   

\textit{Subdivision methods} are recursive approaches that subdivide surfaces into smaller facets and perform intersection tests between the facets of the two surfaces \citep{lasser1986intersection},\   \citep{de1996surface},\   \citep{lin2013affine}.\    These operations are repeated until the computed intersection meets accuracy requirements.\    While subdivision methods can achieve high accuracy in intersection curves,\   they struggle to determine the topology of intersections near singular points.\   

Widely used methods in practice are hybrids of the above approaches.\    One example is the hybrid of subdivision and tracing methods,\    which first identifies all intersecting branches through subdivision and then tracks each intersecting curve \citep{barnhill1990marching},\   \citep{sinha1985exploiting}.\    Another hybrid combines algebraic and tracing methods:\   algebraic methods are used to identify characteristic points,\   and then tracking methods are applied to obtain intersecting curves \citep{krishnan1997efficient}.\    

For more details on Surface/Surface intersection,\   please refer to \citep{hoschek1993fundamentals},\   \citep{patrikalakis2002surface}.\   

In general,\   current research primarily employs algebraic methods to compute these characteristic points of intersection,\   thereby decomposing the intersection curve into several monotonic segments at the characteristic points to determine the topology of the intersection \citep{grandine1997new}\citep{hass2007guaranteed}.\    However,\   few studies focus on the overall topological structure of the intersection,\   such as the number of connected branches and cycles.\   

\subsection{Mapper algorithm and its applications}

The Mapper algorithm was first proposed by Singh et al.\    in 2007 \citep{Singh2007mapper}.\    It aims to perform data visualization and cluster analysis based on topology,\   to extract the global topological features of data.\    The early Mapper algorithm had limitations in practical use,\   such as relying on parameter selections like coverage and overlap ratio.\    To address these issues,\   some studies have proposed adaptive coverage selection strategies based on statistical analysis \citep{carriere2018statistical},\  \citep{belchi2020numerical}.\    In addition,\      scholars have developed Mapper variants based on different clustering algorithms—\    for example,\   applying classic ones like Fuzzy Clustering \citep{bui2020f} and G-Means \citep{alvarado2025g} to Mapper’s clustering step.\    

 Currently,\   the Mapper algorithm has been widely applied in many fields,\   including computational biology \citep{jeitziner2019two},\  medicine \citep{li2015identification},\  
  manufacturing systems \citep{guo2017identification},\   and machine learning \citep{carriere2022statistical}.\    However,\   its application in the CAD field remains scarce.\   
\section{Preliminaries}\label{sec:Preliminaries}

In this section,\   we introduce key concepts used in this paper.\   

\subsection{Mapper}\label{sec:Preliminaries_mapper}

Mapper is  a tool in the field of topological data analysis proposed by Gurjeet Singh et al in 2007 \citep{Singh2007mapper}.\    

The Mapper method extracts the topological structure of a high-dimensional dataset through partial clustering, represented as a simplicial complex.\    In general,\   we usually construct a two-dimensional simplex complex form,\   that is,\   an undirected and unweighted graph $G = \langle V,\   E \rangle$,\   where $V$ stands for nodes and $E$ stands for edges.\    Unless otherwise stated,\   this paper defaults to this form of Mapper's output,\   which we call a \textit{Mapper graph}.\   The Mapper algorithm is inspired by the Reeb graph.\    It can be proven that in the limit,\   the Mapper graph constructed using Mapper is equivalent to the Reeb graph \citep{Munch2015convergence}.\

Given a dataset $X$ with known pairwise distances and a filter function $f:X\rightarrow \mathbb{R}$ ,\  the Mapper algorithm on $X$ computed with the filter function $f$ contains the following steps: 

\begin{enumerate}
    \item {\bfseries Cover construction}.\   Cover the range of values $Y = f(X)$ with a set of overlapping intervals $\{I_s\}_1^S$.\   
    
    \item {\bfseries Preimage clustering}.\    
    For each interval $I_s$,\   the preimage $X_s=f^{-1}(I_s)$ is clustered using a chosen clustering algorithm,\   resulting in clusters $\{X_{s,k}\}$.\   

    \item {\bfseries Graph Construction}.\   A node is created for each cluster 
$X_{s,k}$,\   and edges are added between nodes if their corresponding clusters intersect (i.\   e.\    $X_{s,k} \bigcap X_{t,l} \neq \emptyset$)
\end{enumerate}

\begin{figure}
    
    \subfigure[]{
        \begin{minipage}{0.21\textwidth}
            \centering
            \includegraphics[width=0.9\textwidth]{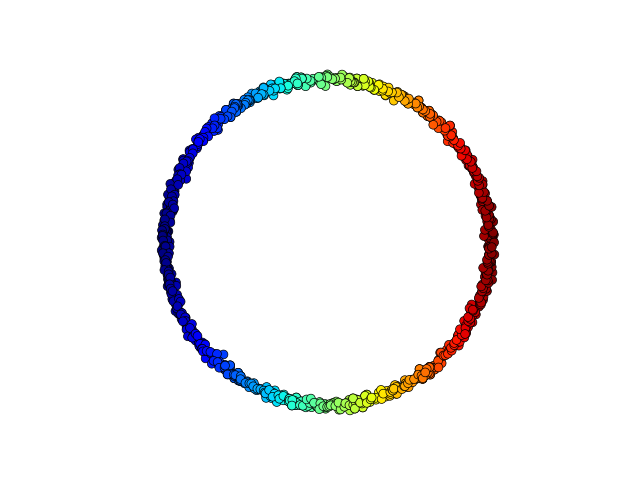}
        \end{minipage}
            }
    \subfigure[]{
        \begin{minipage}{0.23\textwidth}
            \centering
            \includegraphics[width=0.9\textwidth]{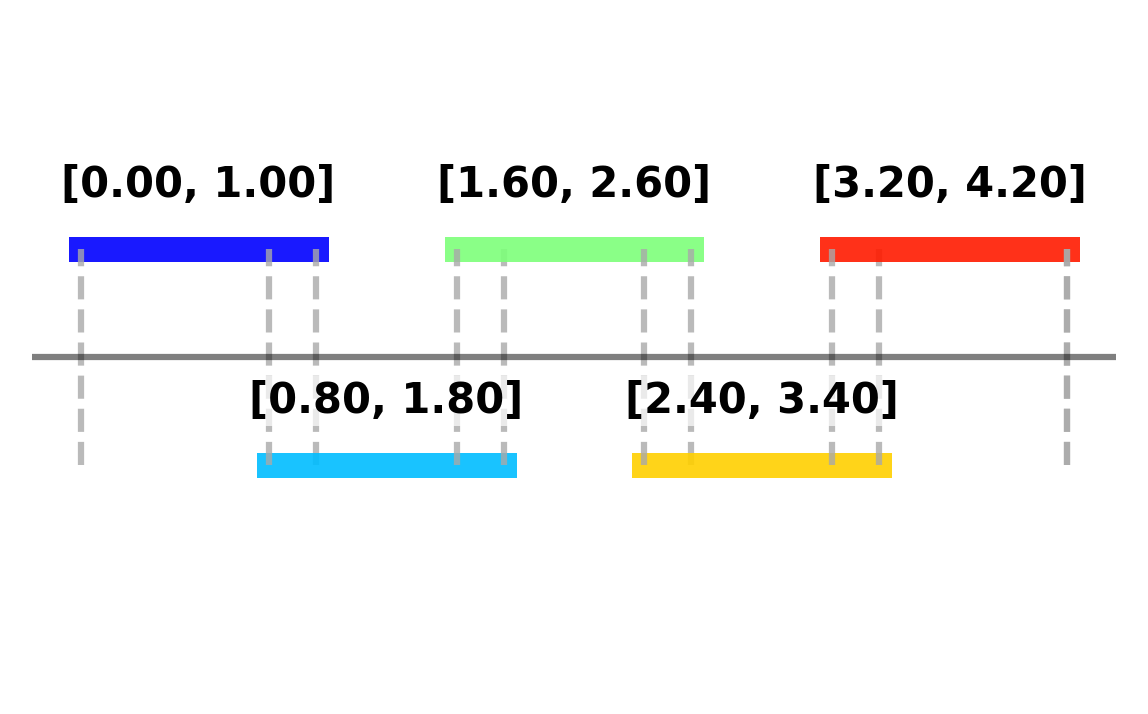}
        \end{minipage}
        }
    \subfigure[]{
        \begin{minipage}{0.25\textwidth}
            \centering
            \includegraphics[width=0.9\textwidth]{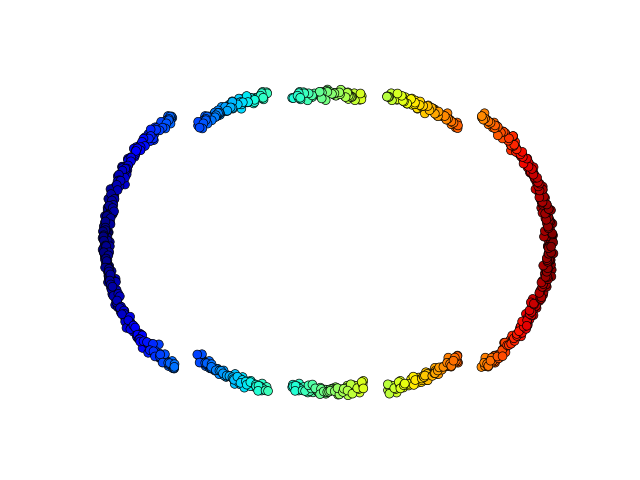}
        \end{minipage}
        }
    \subfigure[]{
        \begin{minipage}{0.2\textwidth}
            \centering
            \includegraphics[width=0.9\textwidth]{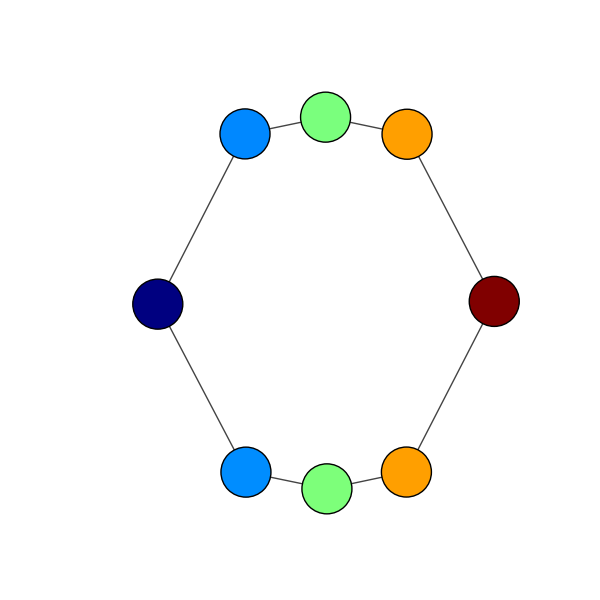}
        \end{minipage}
        }
    \vspace{1em}
    \caption{An example of the Mapper algorithm.\  The data is sampled from a noisy circle, and the filter function is $f (x) = x_1$,\   where $x = (x_1,x_2)$ is a point in $X$.\    (a) The point set $X$ colored by the value of the filter function $f$.\    (b)The range of $f$ is covered with five equal-length intervals,\   with a 20\% overlap of neighboring intervals.\    (c)The DBSCAN algorithm \citep{Ester1996DBSCAN} is used to cluster the preimage for each interval.\    (d)Nodes of the mapper graph are colored by the average filter function value.\   }
    \label{fig:mapper}
    
\end{figure}

Fig. \ref{fig:mapper} illustrates the Mapper algorithm applied to a noisy circle point set $X$.\    The filter function is $f (x) = x_1$,\   where $x = (x_1,x_2)$ is a point in $X$.\    We cover the range of $f$ with five equal-length intervals,\   with a 20\% overlap of neighboring intervals.\    Then for each interval we find its clustering using \textit{DBSCAN (Density-Based Spatial Clustering of Applications with Noise)} algorithm\citep{Ester1996DBSCAN}.\   Finally,\    we treat each cluster as a node in the Mapper graph and add edges between nodes if their corresponding clusters intersect.\   

The choice of filter function,\   cover parameters,\   and clustering algorithm all influence the resulting Mapper graph.\    These parameters must be carefully selected to ensure that the topological features extracted by the algorithm accurately reflect the structure of the point set.\   

\subsection{Surface intersection by subdivision method}

In this section,\   we introduce the general idea of the subdivision method for solving the intersection of B-spline surfaces without involving specific implementations.\   
The core idea of the subdivision method in solving the intersection of B-spline surfaces is to narrow down potential intersection regions through gradual subdivision of the surface parameter domain and the judgment of geometric relationships,\   and finally determine the intersection point sets.\    Its basic steps are as follows.\   

\textbf{Construction of initial bounding boxes and prediction of intersection possibility}.\    For the two B-spline surfaces $B_{1}(u,v)$ and $B_2(s,t)$ involved in the intersection calculation,\   first,\   initial surface patches are divided within their respective parameter domains,\   and an axis-aligned bounding box is constructed for each surface patch.\    By judging whether the bounding boxes of two surface patches intersect,\   we can quickly eliminate the combinations of surface patches that cannot have intersection curves: if the bounding boxes have no intersection,\   the corresponding surface patches must not intersect and can be directly eliminated; if the bounding boxes intersect,\   the surface patches may intersect and need to proceed to the next step of processing.\   

\begin{figure}
    
    \subfigure[]{
        \begin{minipage}{0.21\textwidth}
            \centering
            \includegraphics[width=0.9\textwidth]{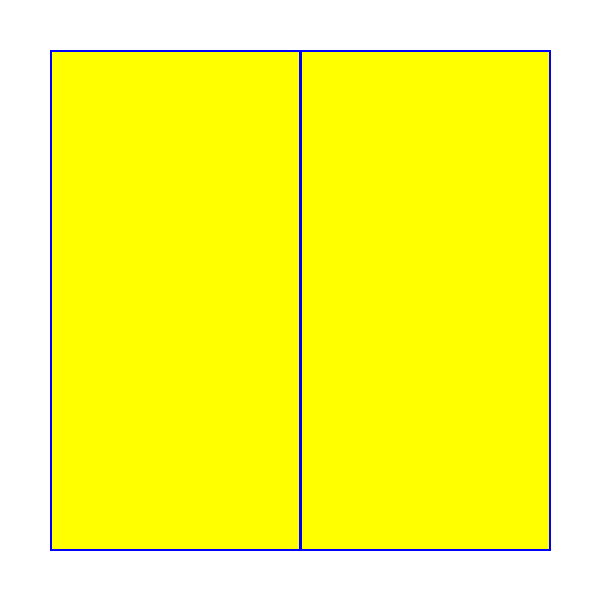}
        \end{minipage}
            }
    \subfigure[]{
        \begin{minipage}{0.21\textwidth}
            \centering
            \includegraphics[width=0.9\textwidth]{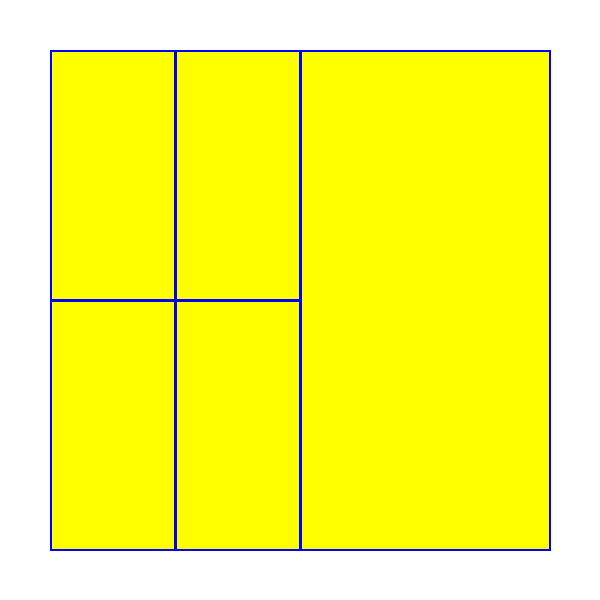}
        \end{minipage}
        }
    \subfigure[]{
        \begin{minipage}{0.21\textwidth}
            \centering
            \includegraphics[width=0.9\textwidth]{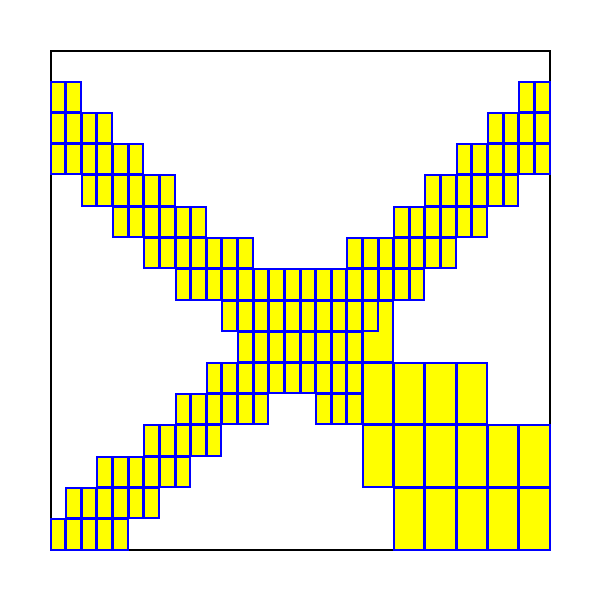}
        \end{minipage}
        }
    \subfigure[]{
        \begin{minipage}{0.21\textwidth}
            \centering
            \includegraphics[width=0.9\textwidth]{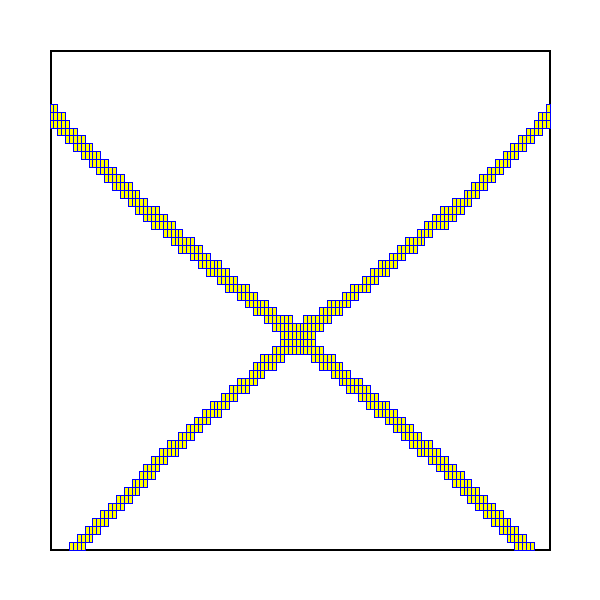}
        \end{minipage}
        }
    \vspace{1em}
    \caption{An example of the subdivision method.\  (a)Initial axis-aligned bounding box on the $(u,v)$ parameter domain.\   Each surface is initially divided into two surface patches within the parameter domain.\  (b)Results after 1 subdivision operation.\   (c)Results after 1000 subdivision operation.\   (d)Results after 4000 subdivision operation.\   }
    \label{fig:subdivision method}
    
\end{figure}

\textbf{Subdivision of parameter domain}.\    For surface patches with intersecting bounding boxes,\   it is necessary to subdivide the corresponding parameter domain rectangles,\   dividing the original rectangles into smaller sub-rectangles,\   each corresponding to a more refined sub-surface patch.\    At the same time,\   a new bounding box is constructed for each sub-surface patch,\   and the above-mentioned intersection possibility prediction process is repeated until the bounding boxes of the sub-surface patches do not intersect (and thus can be eliminated) or the size of the sub-surface patches reaches the preset precision threshold.\   
Fig.\     \ref{fig:subdivision method} illustrates the subdivision process of the $(u,v)$ parameter domain. Each surface is initially divided into two surface patches within the parameter domain.\   Through iterative subdivision and intersection detection,\   the remaining rectangular bounding boxes progressively narrow in the actual intersection.\  

\textbf{Extraction of intersection point set}.\   After iterative subdivision, we obtain two strip-shaped intersection regions on the parameter domains of two B-spline surfaces $B_{1}(u,v)$ and $B_2(s,t)$.\   In fact, they correspond to two sets of rectangles on the respective parameter domains.\   We compute the centroid of each rectangle as a point on the parameter domain where the B-spline surfaces intersect.
In conclusion.\   we get two pieces of intersection point sets 
$P_{1,u,v}$ and $P_{2,s,t}$ on the parameter domains.\   
In order to generate the desirable intersection curves,
	we should understand the topology structure of the strip-shaped point clouds $P_{1,u,v}$ and $P_{2,s,t}$ correctly.

\section{Topology understanding of intersecting curves in parameter domains}

In this paper,\   we use the affine arithmetic-based subdivision method proposed in \citep{lin2013affine} to obtain the intersection point sets $P_{1,u,v}$ and $P_{2,s,t}$ in the parameter domains of the two B-spline surfaces $B_1(u,v)$ and $B_2(s,t)$,\   with their parameter ranges being \([u_s, u_e] \times [v_s, v_e]\) and \([s_s, s_e] \times [t_s, t_e]\).\  
In this section,\   we analyze the topological structure of intersecting curves in the parameter domain based on the Mapper algorithm.\ 

\subsection{Two-step Mapper algorithm}\label{sec:Two-step Mapper}

In Section \ref{sec:Preliminaries_mapper},\   we provide an overview of the basic process of the Mapper algorithm.\    
In this paper,\   We develop a \textit{two-step Mapper} algorithm to extract the topological structure of an input point set $X \subseteq \mathbb{R}^2$.\   This is  outlined in Alg. \ref{alg:two_step_mapper} and explained in the following.   First,\    we construct the initial Mapper graph using a filter function based on principal direction projections.\    Then,\    we subdivide the nodes in the Mapper graph along orthogonal directions to obtain the final Mapper graph.\   Algorithms for higher dimensions can be similarly generalized.\   

\begin{algorithm}[htp]
    \caption{Two-step Mapper algorithm}
    \label{alg:two_step_mapper}
    \begin{algorithmic}[1]
        \REQUIRE A point set $X \subseteq \mathbb{R}^2$,\  clustering parameter $\delta$,\   overlap ratio $\theta_{\text{ov}}$
        \ENSURE Mapper graph $G$ of $X$ 
        \STATE  Construct the initial Mapper graph $G$ 
        \label{step:mapper end}
        \STATE Set $V_{split} = \varnothing$.
        \FOR{each point set \(X_i\) corresponding to the node \(v_i\) in $G$ }
             \STATE  Compute the number of intervals \(S_i\) corresponding to \(X_i\) 
            \IF{$S_i>=2$}
                \STATE Add the node \(v_i\) to $V_{split}$
            \ENDIF
        \ENDFOR
        \STATE Merge nodes in $V_{split}$ that belong to the same connected component in $G$ into a single node.
        \FOR{each point set \(X_i\) corresponding to the node \(v_i\) in  $V_{split}$ }
                \STATE  Construct the  Mapper subgraph $G_i$ corresponding to $X_i$ using $f_{\perp}$.
                \FOR{each point set $X_{nei}$  corresponding to neighboring node $v_{nei}$ of $v_i$ in $G$ }
                    \STATE Merge all nodes in $G_i$ where their corresponding point sets intersect with  $X_{nei}$ into a single node.\   
                \ENDFOR  
                \STATE Replace $v_i$ with all nodes from $G_i$.\   
        \ENDFOR
        \STATE Add new edges to $G$ according to the edge addition rule of the Mapper algorithm.\  
        \RETURN $G$
    \end{algorithmic}
\end{algorithm}

\subsubsection{Generation of the Mapper graph}\label{sec:initial mapper graph}

\begin{figure*}[h]
    
    \centering
     \subfigure[]{
             \label{fig:initial_mapper_0}
         \begin{minipage}{0.3\textwidth}

            \centerline{\includegraphics[width=1\textwidth]{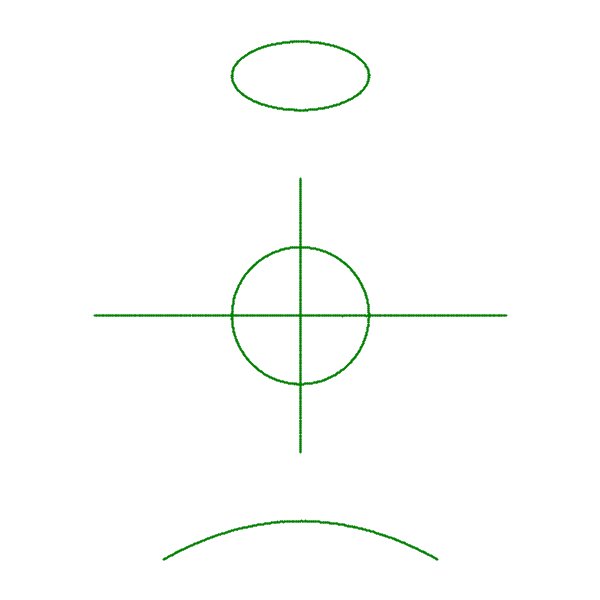}}
                
         \end{minipage}
    }
    \subfigure[]{
        \label{fig:initial_mapper_1}
         \begin{minipage}{0.3\textwidth}
            \centerline{\includegraphics[width=1\textwidth]{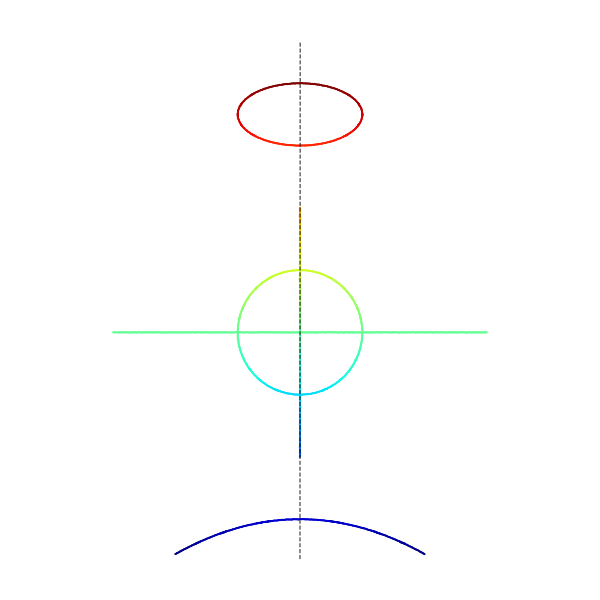}}
                
         \end{minipage}
    }
     \subfigure[]{
        \label{fig:initial_mapper_2}
         \begin{minipage}{0.3\textwidth}
            \centerline{\includegraphics[width=1\textwidth]{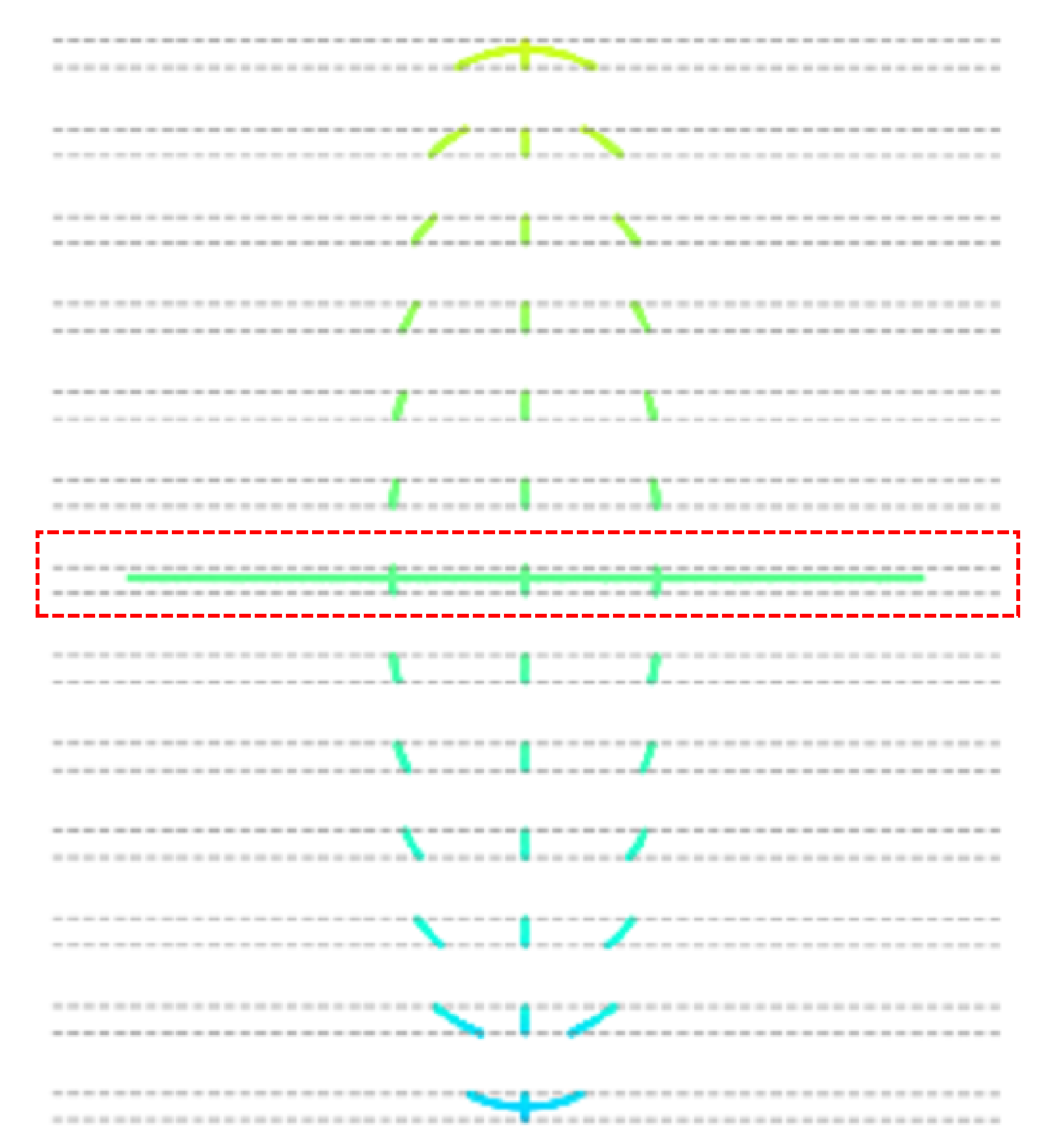}}
                
         \end{minipage}
    }
    
    \subfigure[]{
             \label{fig:initial_mapper_3}
         \begin{minipage}{0.3\textwidth}

            \centerline{\includegraphics[width=1\textwidth]{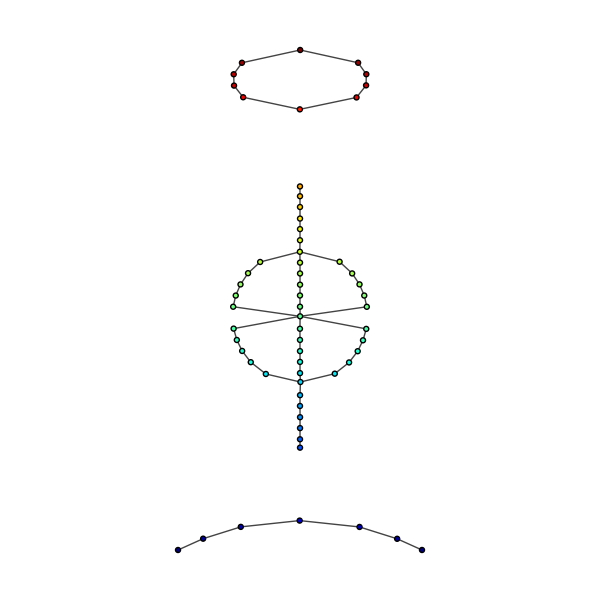}}
                
         \end{minipage}
    }
    \subfigure[]{
             \label{fig:two_step_mapper_1}
         \begin{minipage}{0.3\textwidth}

            \centerline{\includegraphics[width=1\textwidth]{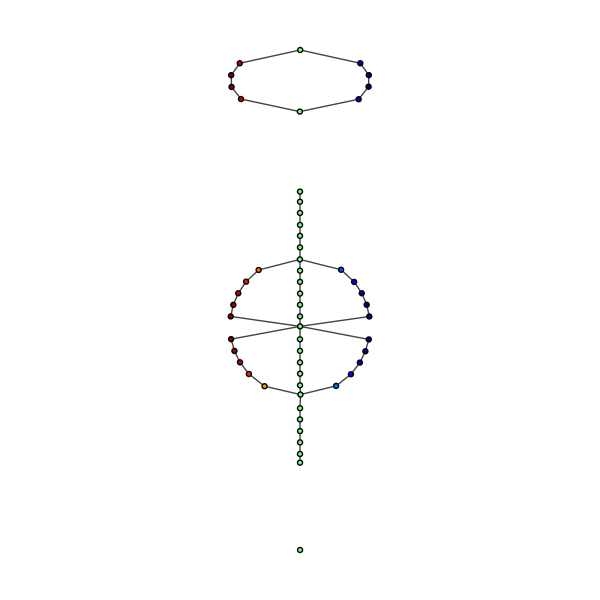}}
                
         \end{minipage}
    }
    \subfigure[]{
             \label{fig:two_step_mapper_2}
         \begin{minipage}{0.3\textwidth}

            \centerline{\includegraphics[width=1\textwidth]{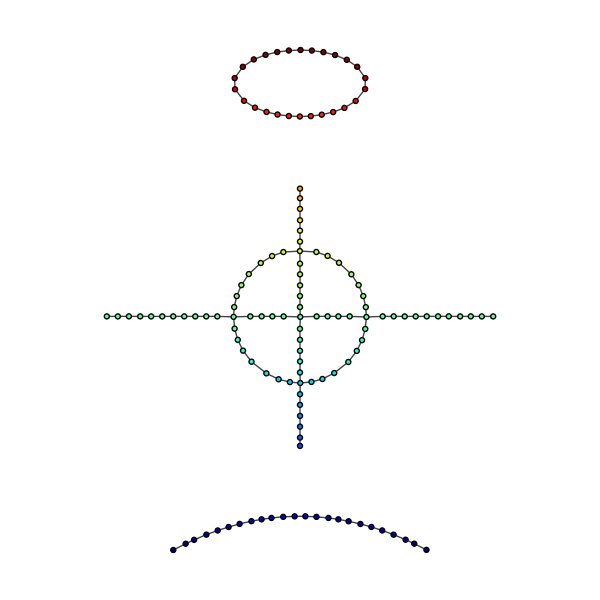}}
                
         \end{minipage}
    }
    
     \caption{Generation of the initial Mapper graph,\   $\theta_{\text{ov}}=0.2$.\   (a)Input point set $X$.\   (b)$X$ is colored according to the values of the filter function $f$.\   The black dashed line passes through the center of $X$,\    indicating the principal direction of $X$.\   (c)Cover construction.\  Only a subset of intervals is visualized for clarity.\   Points corresponding to the same interval are enclosed between dashed lines,\   and the preimages of different intervals are translated for better visual separation (d)Initial Mapper graph.\   (e)The result of the node merging operation.\   (f)The final Mapper graph.}
     \label{fig:initial_mapper}
\end{figure*}

\begin{figure}
    \centering
    \includegraphics[width=0.9\linewidth]{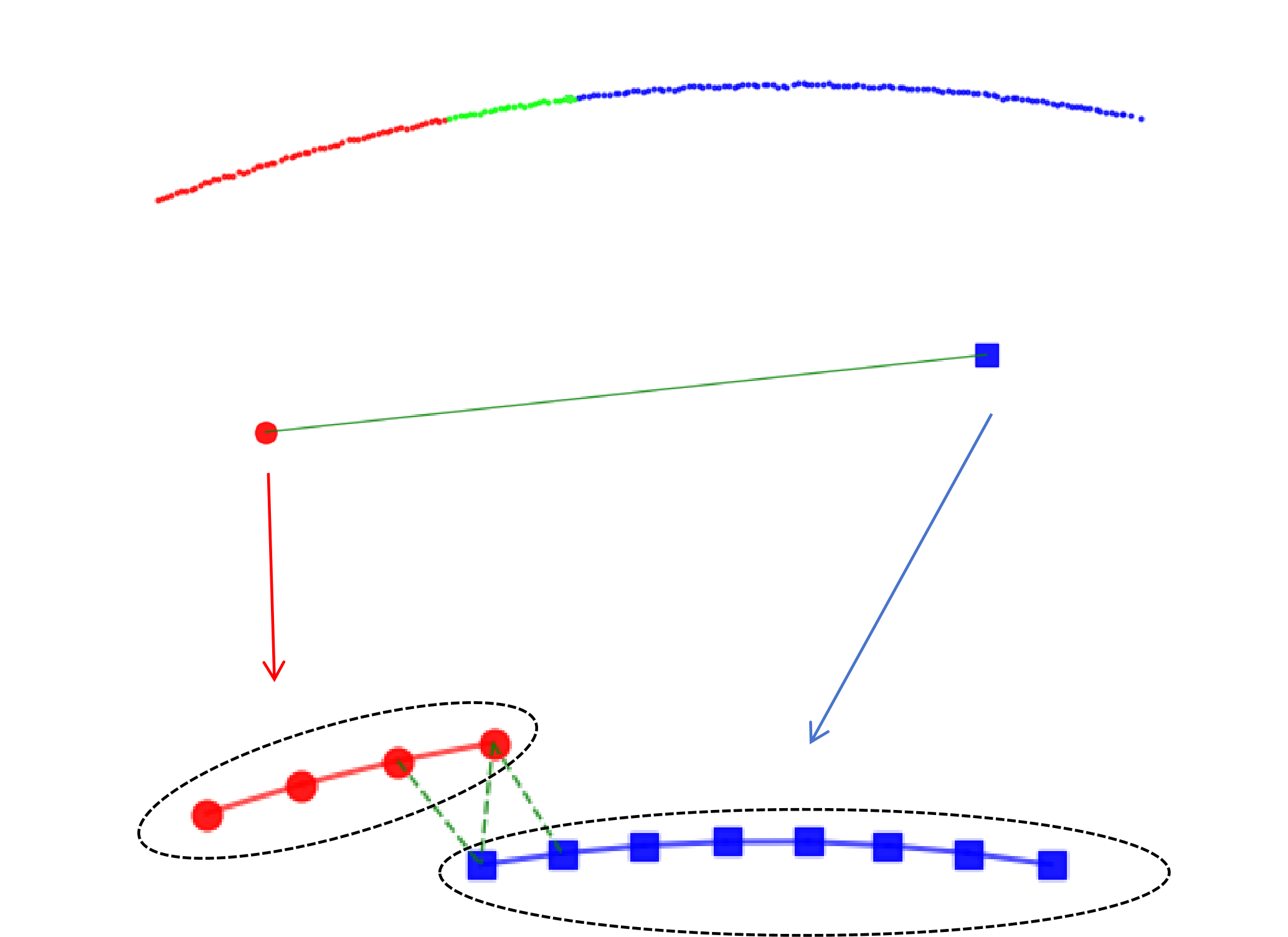}
    \vspace{1em}
    \caption{An illustration of additional cross-graph edges introduced by independent subdivision of adjacent Mapper nodes.\   
Top:\   data elements corresponding to two adjacent nodes,\   where node-specific elements are shown in red and blue,\   and their shared elements are shown in green.\   
Middle:\   the corresponding two adjacent Mapper nodes and their connecting edge.\  
Bottom:\   Mapper subgraphs generated independently from the two nodes.\  
Due to inconsistent partitioning of shared elements, the edge addition rule based on whether corresponding element sets intersect introduced additional cross-graph edges (green dashed lines).}
    \label{fig:mapper_split}
\end{figure}

This section follows the work of Carrière et al.\ \citep{carriere2018statistical},\   who investigated the convergence of the Mapper graph to its continuous analogue,\   namely the Reeb graph.\   
They employed extended persistence and its associated metric,\   the bottleneck distance \citep{edelsbrunner2010computational},\   to quantitatively characterize the topological similarity between the Mapper graph and the corresponding Reeb graph.\   
Their analysis provides parameter selection criteria that guarantee the bottleneck distance between the extended persistence diagrams of the Mapper graph and the Reeb graph does not exceed a parameter-dependent upper bound.\   
To ensure the topological correctness of the Mapper graph,\   we adopt the Mapper algorithm and parameter selection strategy proposed by Carrière et al.\   
The specific implementation details used in this paper are described below.\

\textbf{Filter function selection}.\   
To obtain general theoretical guarantees, 
	the filter function should be Morse-type.\   
A large class of Morse-type functions commonly used in Mapper constructions are Lipschitz continuous,\   i.e.,\   there exists a constant $c>0$ such that the function $f:X\rightarrow \mathbb{R}$ satisfies
\begin{equation}
    |f(x)-f(x^{\prime})|\leq c\|x-x^{\prime}\|.
\end{equation}

In this paper,\   we employ linear projections as filter functions,\   for which the Lipschitz constant satisfies $c=1$.\   
When constructing the initial Mapper graph,\   we apply principal component analysis (PCA) \citep{jolliffe1990principal} to compute the principal directions of $X$,\   and adopt the projection of the data onto the first principal direction as the filter function.\   
Specifically,\   we first compute the centroid $x_{c}$ of $X$,\   
\begin{equation}\label{equ:center_of_x}
    x_c= \frac{1}{m}\sum_{i=1}^{m}x_i.
\end{equation}
Next,\   we compute the covariance matrix $\Sigma =(\sigma_{i,j})$ of the centralized point set $X$,\   
\begin{equation}\label{equ:covariance_matrix}
    \sigma_{i,j} = \langle x_i -x_c,\  x_j-x_c \rangle,\   \text{where } x_i,\  x_j \in X.
\end{equation}
The eigenvector corresponding to the largest eigenvalue of $\Sigma$ is denoted by $w_p$ and defines the \textit{principal direction}.\   
The filter function $f$ is then defined as the projection of a data point $x$ onto $w_p$:
\begin{equation}\label{equ:projection_filter_function}
    f(x) = \langle x -x_c,\  w_p \rangle.
\end{equation}

\textbf{Clustering method and parameter calculation}.\   
To guarantee the topological correctness of the Mapper graph,\   the parameters used to construct the cover are intrinsically related to the clustering parameters.\   
Therefore, 
we first introduce the clustering method and parameter computation before describing the cover construction.\   

In the clustering step,\   we construct a $\delta$-neighborhood graph based on $X$,\   where an edge is drawn between two distinct points if their Euclidean distance is less than $\delta$.\   
The connected components of the preimage of the filter function $f$ induced by this $\delta$-neighborhood graph are then taken as the clustering results.\   

The parameter $\delta$ should satisfy the following conditions:
\begin{equation}\label{equ:epsilon2}
    4 d_{\mathrm{H}}\left(\mathcal{M}, X\right) \leq \delta,
\end{equation}
\begin{equation}\label{equ:epsilon1}
     \delta \leq \frac{1}{4} \min \{rch, \rho\},
\end{equation}
where $rch$ and $\rho$ denote the reach and convexity radius of $\mathcal{M}$,\  respectively,\   and $d_{\mathrm{H}}$ denotes the Hausdorff distance.\   
Here,the reach $rch$  is defined as the supremum of all $r>0$ such that every point $x$ with the Euclidean distance $\mathrm{dist}(x,\mathcal{M})<r$ admits a unique nearest point in $\mathcal{M}$.
The convexity radius $\rho$ is the largest $r>0$ such that for every $p\in\mathcal{M}$, the geodesic ball
$B_{\mathcal{M}}(p,r)=\{q\in\mathcal{M}: d_{\mathcal{M}}(p,q)<r\}$
is geodesically convex.

Regarding condition~\eqref{equ:epsilon2},\   since each intersection point is taken as the center of its corresponding bounding box,\   an upper bound for $d_{\mathrm{H}}\left(\mathcal{M}, X\right)$ is given by half the length of the bounding box diagonal.\   
Accordingly,\   we select twice the diagonal length of the bounding box as the value of $\delta$.\   

For condition~\eqref{equ:epsilon1},\   if the intersection does not contain singular points (i.e.,\   points of tangential discontinuity),\   the numerical accuracy of standard surface/surface intersection algorithms is generally sufficient,\   and the above choice of $\delta$ naturally satisfies this constraint.\   
However,\   if the intersection contains singular points,\   the manifold induced by the corresponding bounding box set may fail to possess a positive reach radius $rch$ and convexity radius $\rho$.\   
This leads to excessive connectivity among nodes in the Mapper graph in the vicinity of these singular points.\     We will address this issue later.\   

\textbf{Cover construction}.\   
Given the filter function $f:X\rightarrow \mathbb{R}$,\   we construct a uniform cover of $f(X)$ such that no more than two intervals overlap at any point.\   
Specifically,\   we use $S$ open intervals $\{I_s\}_{s=1}^{S}$ of equal length $l$.\   
The overlap ratio $\theta_{\text{ov}}$ between adjacent intervals is defined as
\begin{equation}\label{equ:r_over}
    0< \theta_{\text{ov}} =\frac{\ell(I_s\cap I_{s+1})}{l}<\frac{1}{2},
\end{equation}
where $\ell(\cdot)$ denotes the Lebesgue measure on $\mathbb{R}$.\   

The interval length $l$ should satisfy the following condition related to the clustering parameter $\delta$ and the filter function $f$:
\begin{equation}\label{equ:comput_interval_len}
    l > l_0= \frac{\mathop{sup}\limits_{\|x_i-x_j\|<\delta,\  x_i\in X,\  x_j\in X} |f(x_i)-f(x_j)|}{\theta_{\text{ov}}}.
\end{equation}

Since the number of intervals $S$ and the interval length $l$ satisfy,
\begin{equation}\label{equ:interval_num_len}
    Sl-(S-1)\theta_{\text{ov}}l = \mathop{sup}\limits_{x_i\in X} f(x_i) - \mathop{inf}\limits_{x_i\in X} f(x_i),
\end{equation}
To ensure condition~\eqref{equ:comput_interval_len},\   we compute the number of intervals $S$ as

\begin{equation}\label{equ:compute_interval_num}
    S = \left\lfloor \frac{\mathop{sup}\limits_{x_i\in X} f(x_i) - \mathop{inf}\limits_{x_i\in X} f(x_i) - \theta_{\text{ov}}l'}{(1-\theta_{\text{ov}})l'} \right\rfloor,
\end{equation}
where $\lfloor \cdot \rfloor$ denotes the floor operator and $l'=(1+\alpha)l_0$.\   
In this paper,\   we set $\alpha=0.001$ to ensure $l'>l_0$.\

Fig.\ \ref{fig:initial_mapper} illustrates an example of constructing the initial Mapper graph $G$.\   
To better reflect the practical setting considered in this paper,\   we uniformly sample points from several planar curves using a fixed arc-length step size $sample = 0.02$  
and add random offset noise with magnitude in the range $[0,\ 0.01]$ to each sampled point,\   yielding the point cloud $X$.\   
According to the parameter selection criterion in Eq. ~\eqref{equ:epsilon1},\   we therefore set
\[
\delta = 4\left(\epsilon + \frac{sample}{2}\right)
\]
in this example.\   

We then compute the centroid and the principal direction of $X$ using PCA,\   and evaluate the filter function $f$ accordingly.\   
In Fig.\ \ref{fig:initial_mapper_1},\   the points in $X$ are colored according to their corresponding filter function values.\   
The black dashed line indicates the line segment passing through the centroid of $X$ and aligned with its principal direction.\  
In Fig.\ \ref{fig:initial_mapper_2},\   we employ $46$ intervals to uniformly cover the range of the filter function $f$.\   
Finally,\   Fig.\ \ref{fig:initial_mapper_3} presents the resulting initial Mapper graph constructed from these settings.\

\subsubsection{Subdivision in the orthogonal direction}

In Fig.\ \ref{fig:initial_mapper_2},\   the points sampled from the middle curve segment in the example exhibit a very narrow range of values under the filter function $f$.\   
As a result,\   these points are assigned to the same interval,\   and are consequently aggregated into a single node in the Mapper graph,\   as indicated by the dashed box.\   
Such aggregation weakens the ability of the Mapper graph to capture the geometric structure of the underlying data.\   

To handle this issue,\   we analyze the initial Mapper graph $G$ in the direction orthogonal to the principal direction vector $w_p$ in Eq. \eqref{equ:projection_filter_function}.\   The core idea is to use the Mapper algorithm to split nodes with excessively large projection spans in the orthogonal direction into smaller nodes.\   

Let $w_{\perp}$ be the unit vector orthogonal to the principal direction vector $w_p$,\   i.e.\    $\langle w_{\perp},\  w_p\rangle = 0$.\   
We define a new filter function $f_{\perp}$  as the projection of the data point $x$ onto $w_{\perp}$ as follows:
\begin{equation} \label{equ:orthogonal projection filter function}
    f_{\perp}(x) = \langle x_i -x_c,\  w_{\perp} \rangle
\end{equation}
where $x_c$ is the center of $X$  calculated by Eq. \eqref{equ:center_of_x}.\

Let \(X_i\) denote the set of points corresponding to each node \(v_i\) in $G$. In Eq. \eqref{equ:comput_interval_len} and \eqref{equ:compute_interval_num} , we substitute $f_{\perp}$ for $f$ and \(X_i\) for $X$ to compute the number of intervals \(S_i\) that \(X_i\) can partition in the orthogonal direction.\   

We denote the set of nodes awaiting splitting as $V_{split} = \{v\in G,S_i\geq 2 \}$.\   
The straightforward idea is to use $f_{\perp}$ instead of $f$ in the Mapper algorithm for each node $v$ in $V_{split}$,\    obtaining the subgraph $G_i$ and replace each such $v_i$ with all nodes in $G_i$ and add edges between different subgraph nodes based on the intersection relationships of corresponding clusters to achieve subdivision.\    

However,\    if two adjacent nodes both require to be subdivided,\    this method may introduce additional edges because the classification results for elements in their overlapping region might differ across their subgraphs.\    
However,\   when two adjacent nodes in the Mapper graph are both subject to subdivision,\   this strategy may introduce additional edges.\   
The reason is that elements lying in the overlapping region of the two nodes may be classified differently in their respective subdivided subgraphs.\   

Fig.\ \ref{fig:mapper_split} provides an illustrative example.\   
The two original nodes share a common subset of $X$.\   However,these elements  is separated into different clusters during the generation of the Mapper subgraph.\   
Consequently,\   additional cross-graph edges are introduced.\

Therefore,\    to achieve effective subdivision while avoiding the generation of additional edges,\    we designed the following process for handling $V_{split}$.\   

First,\   we merge all adjacent nodes in $V_{split}$ into a single node.\    If multiple nodes in $V_{split}$ are mutually adjacent,\    we simultaneously merge them into a single node in $G$.\    For brevity,\    we continue to denote the modified graph and the set of nodes to be split as $G$ and $V_{split}$,\    respectively.\   

Next,\   for each node $v_i$ in $V_{split}$,\   we use $f_{\perp}$ instead of $f$ in the Mapper algorithm in section \ref{sec:initial mapper graph} to obtain the Mapper graph $G_i$ for $X_i$.\   \   

It should be noted that the set of common points  between node \(v_i\) and one of its neighboring nodes \(v_j\) may be assigned to multiple nodes in \(G_i\).\    In such cases, directly replacing \(v_i\) with  all nodes from \(G_i\) may introduce redundant edges in the modified graph,as shown in Fig. \ref{fig:mapper_split}.\    Therefore, we iterate through each neighboring node \(v_j\) of \(v_i\) and merge all nodes in graph $G_i$ that share common points with \(v_j\) into a single node.\   Subsequently, we modify $G$ by replacing \(v_i\) with all nodes in \(G_i\).

Finally,\   we add new edges to modified graph $G$ between nodes following the Mapper edge addition rule ,\   i.\   e.\    edges exist if the intersection of the point sets corresponding to two nodes is non-empty.\

Fig. \ref{fig:two_step_mapper_1} illustrates the result of the node merging operation.\   Specifically,\   the nodes in the lowest connected subgraph of $G$ are merged into a single node.\   
Fig. \ref{fig:two_step_mapper_2} shows the final modified Mapper graph. Compared to the initial Mapper graph in Fig. \ref{fig:initial_mapper_3},\    the modified Mapper graph contains more nodes in the orthogonal direction, thereby better reflecting the structure of the input point set $X$.\  

\subsection{Partition of intersection set based on  the  Mapper graph}

Using the two-step Mapper algorithm,\   we can generate the Mapper graph $G$ of the intersection point set $P_{1,u,v}$.\   
Next,\    we partition $P_{1,u,v}$ based on the characteristic nodes of the  Mapper graph.\     

We define two types of characteristic nodes in the Mapper graph:\    \textit{boundary nodes} and \textit{singular nodes}.\   They correspond respectively to the boundary points and singular points of the intersection point set.\   

Let the parameter range of the surface $B_1(u,v)$ be \([u_s, u_e] \times [v_s, v_e]\).\   A \textit{boundary point} is an intersection point on the boundary lines:\ 
\begin{equation*}
    l = \left\{ (u, v) \in \mathbb{R}^2 \mid u = u_s \text{ or } u = u_e \text{ or } v = v_s \text{ or } v = v_e \right\}
\end{equation*}
Due to the overestimation property of the subdivision method,\    we need to perform dilation on the boundary point set.\    Specifically,\    we regard the points with a distance less than \(\) the clustering parameter $\delta$ to the boundary lines as the approximate boundary point set:\   

\begin{multline}
P_{b} = \{ (u, v) \in P_{1,u,v} \mid 
u < u_s + \delta, 
\text{or } u > s_e - \delta, \\
\text{or } v < v_s + \delta, 
\text{or } v > v_e - \delta  \}
\end{multline}
Then we define a node $v_{bou}$ as a \textit{boundary node} if its corresponding point set intersects with the approximate boundary point set $P_b$.\   

A \textit{singular point} is a point where the curve does not have a unique tangent.\   For the intersection of two B-spline surfaces,\    such points are typically located within the cross-type regions of the intersection point set.\   These cross-type regions correspond to nodes with a degree greater than 2 in graph $G$.\   Therefore, we define a node $v_{sin}$ in the Mapper graph $G$ with a degree greater than 2 as a \textit {singular node}.\   

Then, we remove all boundary nodes and singular nodes from the Mapper graph $G$.\   
Finally,\    we partition $P_{1,u,v}$ based on the connection relationships between nodes in $G$.\      Specifically,\   we group all nodes in $G$ that are connected by a walk—i.e.,   nodes forming a connected component of $G$—\   into one group.\   The point sets corresponding to these nodes form a subset of $P_{1,u,v}$.\   In this way,\   we partition $P_{1,u,v}$ into several subsets.\   Since each connected component of the modified graph $G$ is either a path,\    a cycle,\    or an isolated node,\    where,\   
\begin{itemize}
\item Each path corresponds to a simple open curve segment in the intersection set;
\item Each cycle corresponds to a simple closed curve segment;
\item Each isolated node corresponds to an isolated intersection point.
\end{itemize}
Thus,\   through such grouping,\   we can partition $P_{1,u,v}$ into several simple segments.\

\begin{figure*}[htbp]
    \centering
    \subfigure[Two intersecting cylindrical surfaces]{
        \label{fig:partition_1}
         \begin{minipage}{0.33\textwidth}
            \centerline{\includegraphics[width=1.0\textwidth]{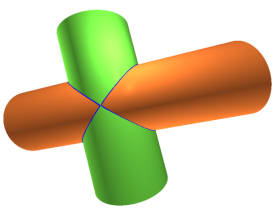}}
                
         \end{minipage}
    }
     \subfigure[The intersection point set $P_{1,u,v}$ ]{
        \label{fig:partition_2}
         \begin{minipage}{0.3\textwidth}
            \centerline{\includegraphics[width=0.9\textwidth]{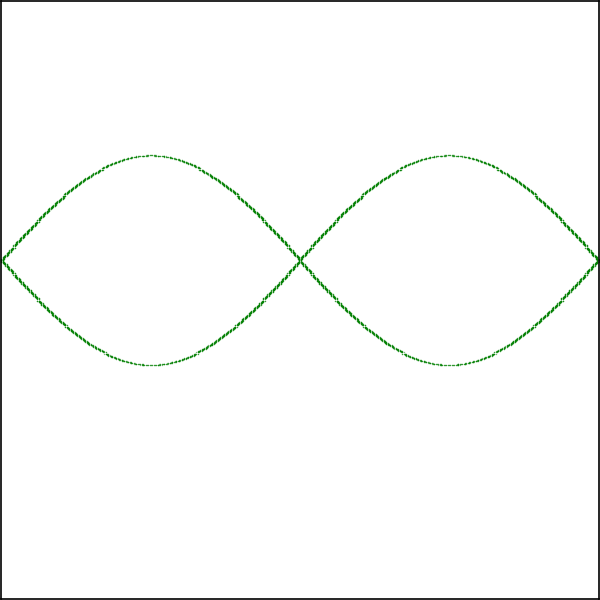}}
                
         \end{minipage}
    }
    \subfigure[The Mapper graph of $P_{1,u,v}$ ]{
             \label{fig:partition_3}
         \begin{minipage}{0.3\textwidth}
        \centerline{\includegraphics[width=0.9\textwidth]{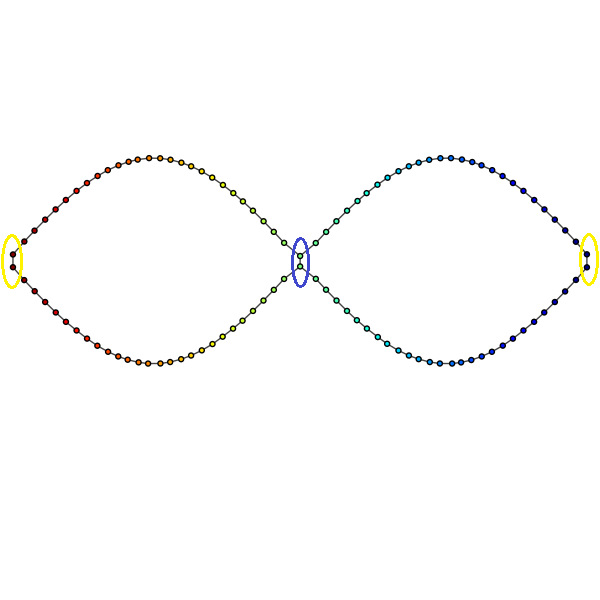}}
                
         \end{minipage}
    }
    
     \subfigure[Mapper graph after removing characteristic nodes]{
        \label{fig:partition_4}
         \begin{minipage}{0.3\textwidth}
            \centerline{\includegraphics[width=0.9\textwidth]{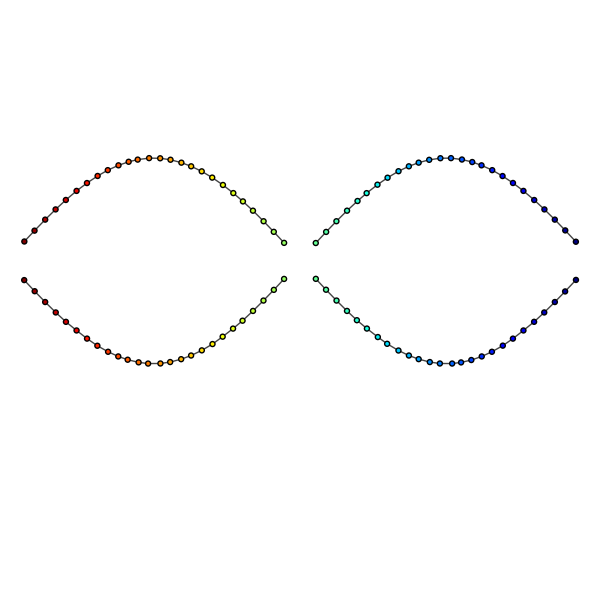}} 
         \end{minipage}
    }
     \subfigure[The partitioning result of $P_{1,u,v}$]{
        \label{fig:partition_5}
         \begin{minipage}{0.3\textwidth}
            \centerline{\includegraphics[width=0.9\textwidth]{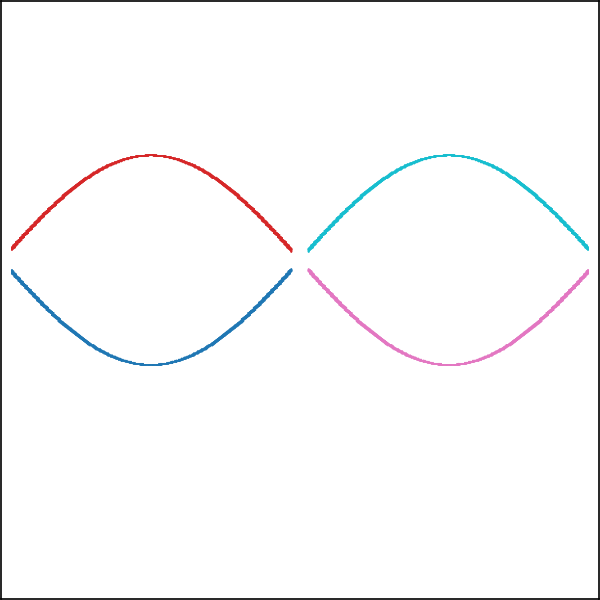}}
                
         \end{minipage}
    }
    \subfigure[Corresponding results of $P_{2,s,t}$]{
        \label{fig:partition_6}
         \begin{minipage}{0.3\textwidth}
            \centerline{\includegraphics[width=0.9\textwidth]{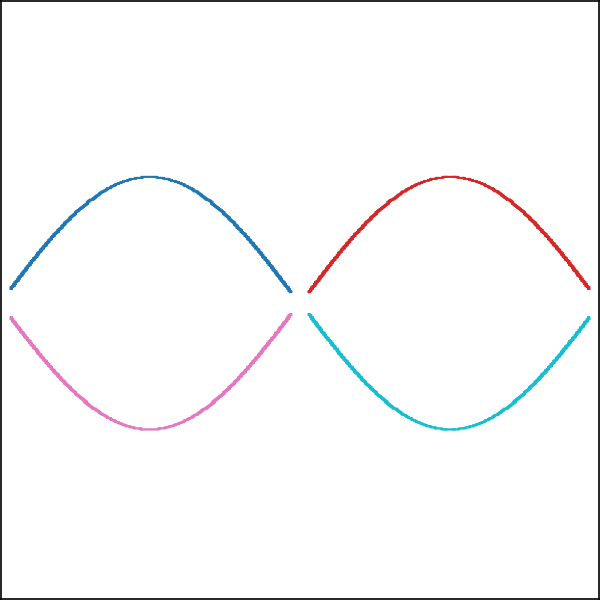}}             
         \end{minipage}
    } 
     \caption{An example of intersecting set partition based on the Mapper graph.\   }
     \label{fig:partition}
\end{figure*}

Fig. \ref{fig:partition} illustrates the process of intersection set partition based on the Mapper graph.\     In Fig. \ref{fig:partition_3},\     yellow and blue circles mark boundary nodes and singular nodes,\     respectively.\     After removing these two types of characteristic nodes,\     the Mapper graph shown in Fig. \ref{fig:partition_4} contains four connected components.\     Based on this,\     we can partition the intersection set $P_{1,u,v}$ into four mutually exclusive subsets and visualize each subset using distinct colors in Fig. \ref{fig:partition_5}.\    

\subsection{Correspondence between different parameter domains.\   }   
Unlike algebraic methods,\   subdivision methods inherently preserve correspondence between intersection results across different surface parameter domains.\   Specifically,\   the subdivision method obtains intersection point sets by recursively detecting bounding box intersections in different parameter domains.\   In the resulting set $P_{1,u,v}$,\   each point’s correspondence to points in $P_{2,s,t}$ is determined by the intersection relationships between their respective bounding boxes.\   

Fig. \ref{fig:partition_6} illustrates the corresponding result for $P_{2,s,t}$,\   where matching partitioned subsets between $P_{1,u,v}$ and $P_{2,s,t}$ are colored identically.\

\section{Experiments and discussions}\label{sec:Experiments and discussions}

\begin{table}[htbp]
    \centering
    \caption{Runtime for constructing the Mapper graph ,\   where ``Initial'' is the time for constructing the initial Mapper graph,\   ``Refinement'' is the time for graph subdivision,\   and ``Total'' is the overall runtime.}
    \label{tab:exp_quantitative}
    \setlength{\tabcolsep}{3pt} 
    \begin{tabular}{c c c c c}
        \toprule  
        Example & \makecell{Number of\\ Points}&  Initial(s) & Subdivision (s) & Total (s) \\
        \midrule  
        Example 1       & 3214    & 0.1340 & 0.1573 & 0.2913 \\
         Example 2       & 2176   &  0.0945 & 0.4427 & 0.5372 \\
         Example 3       & 1851    & 0.1278 & 0.1242 & 0.2519 \\
         Example 4       & 1902   &   0.0901 & 0.0661 & 0.1562 \\
         Example 5       & 3318    &  0.1238 & 0.1677 & 0.2915 \\
         Example 6       & 2650    &   0.1262 & 0.0830 & 0.2092 \\
         Example 7      & 5352   &  0.1551 & 0.1800 & 0.3350  \\
        Example 8       & 4772        &   0.1125 & 0.3022 & 0.4148 \\
         Example 9       & 6284       &   0.1673 & 0.3527 & 0.5200 \\
        \bottomrule  
    \end{tabular}
\end{table}

\begin{figure*}[hbp]
    \centering
    \subfigure[The intersection point set $P_{1,u,v}$ ]{
        \label{fig:exp_overlap_1}
         \begin{minipage}{0.15\textwidth}
            \centerline{\includegraphics[width=0.9\textwidth]{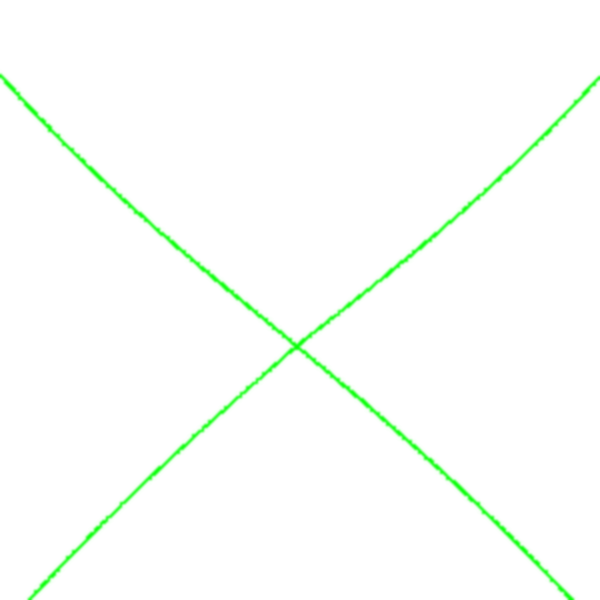}}
                
         \end{minipage}
    }
     \subfigure[$\theta_{\text{ov}}=0.1$]{
        \label{fig:exp_overlap_2}
         \begin{minipage}{0.15\textwidth}
            \centerline{\includegraphics[width=0.9\textwidth]{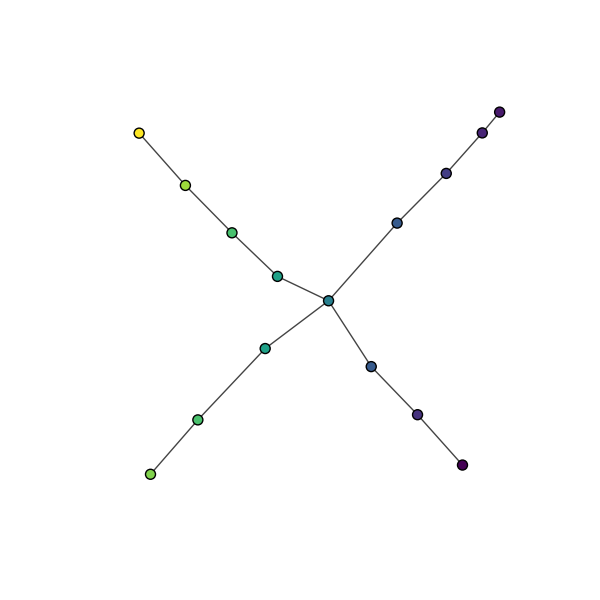}}
                
         \end{minipage}
    }
    \subfigure[$\theta_{\text{ov}}=0.2$]{
             \label{fig:exp_overlap_3}
         \begin{minipage}{0.15\textwidth}
        \centerline{\includegraphics[width=0.9\textwidth]{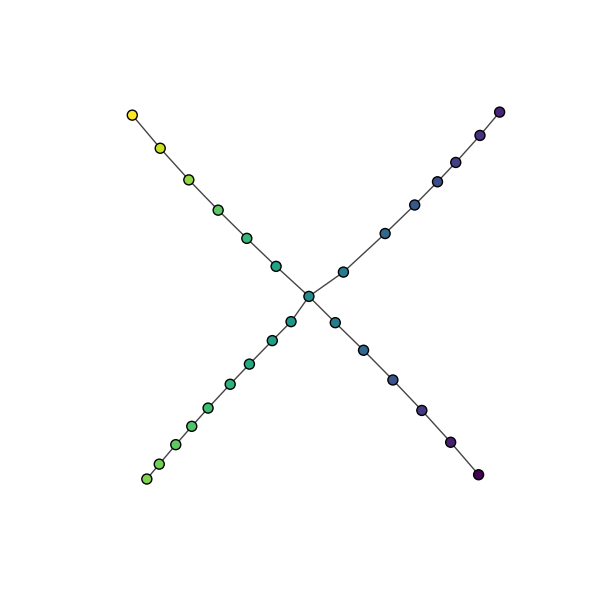}}
                
         \end{minipage}
    }
     \subfigure[$\theta_{\text{ov}}=0.3$]{
        \label{fig:exp_overlap_4}
         \begin{minipage}{0.15\textwidth}
            \centerline{\includegraphics[width=0.9\textwidth]{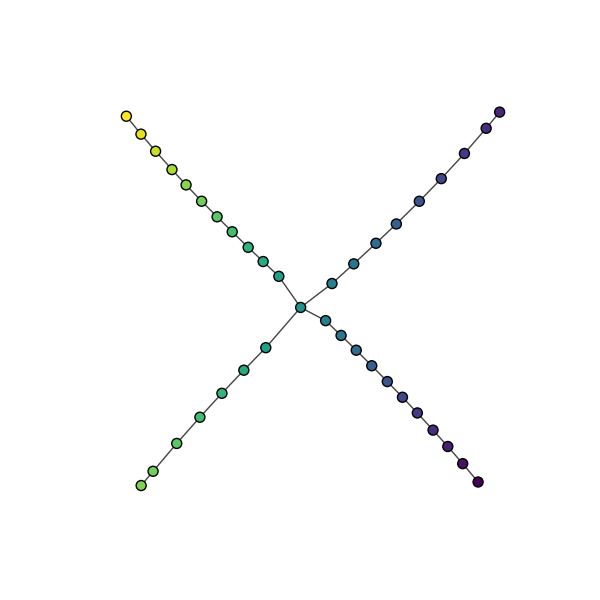}} 
         \end{minipage}
    }
     \subfigure[$\theta_{\text{ov}}=0.4$]{
        \label{fig:exp_overlap_5}
         \begin{minipage}{0.15\textwidth}
            \centerline{\includegraphics[width=0.9\textwidth]{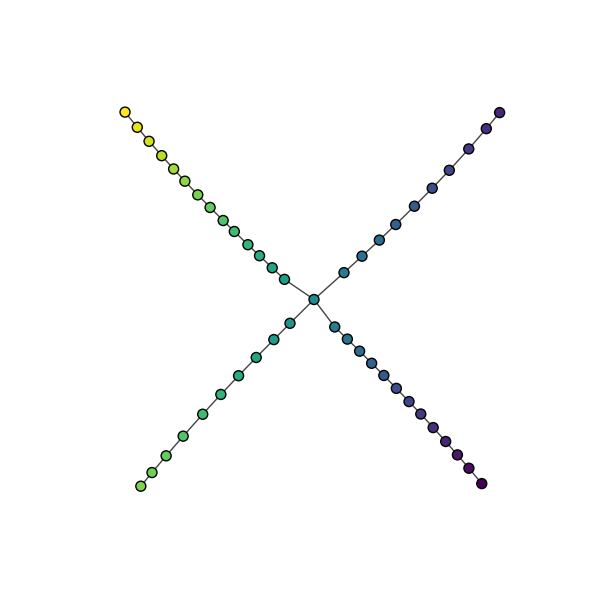}}
                
         \end{minipage}
    }

     \caption{An example of the impact of changing the overlap ratio parameter $\theta_{\text{ov}}$.\   }
     \label{fig:exp_overlap}
\end{figure*}
All experiments in this paper were implemented on a PC with a Intel Core Ultra 7 155H 3.80GHz CPU and 32GB RAM.\   
We use python's Giotto-tda library \citep{tauzin2021giotto} to compute the Mapper graph.\   
In our experiments,\   we will show that the algorithm can successfully understand the topology of intersection.\   

\subsection{Parameter selection for the Two-step Mapper algorithm}

The overlap ratio $\theta_{\text{ov}}$ is a key parameter in the Mapper algorithm. This section analyzes its influence on the resulting graph structure.

The overlap ratio $\theta_{\text{ov}}$ controls the degree of overlap between two adjacent intervals when constructing the cover. 
According to Eq.~\ref{equ:comput_interval_len}, the interval length $l$ is proportional to $1/\theta_{\text{ov}}$. 
Consequently, increasing $\theta_{\text{ov}}$ leads to shorter intervals and a larger number of nodes in the resulting Mapper graph. This produces a more compact and detailed graph representation, but at the expense of increased computational cost.

Figure~\ref{fig:exp_overlap} illustrates the effect of varying $\theta_{\text{ov}}$. 
When $\theta_{\text{ov}} = 0.1$, the graph is sparsely connected due to the large interval length. 
When $\theta_{\text{ov}} = 0.2$, the node distribution becomes significantly denser and more informative. 
When $\theta_{\text{ov}}$ is further increased to $0.4$, the graph continues to become denser; 
however, the growth trend becomes less pronounced. 
This behavior is consistent with the relationship between $l$ and $\theta_{\text{ov}}$. 
To balance the expressive capability of the Mapper representation and computational efficiency, we set $\theta_{\text{ov}} = 0.2$ in the following experiments.

\subsection{Experimental results}
In this section,\    we apply the proposed algorithm to a series of surface/surface intersections  with complex topology to validate its effectiveness.\   

Table \ref{tab:exp_quantitative} reports the runtime statistics for constructing the initial Mapper graph,\   performing Mapper graph subdivision,\   and the total runtime.\   
It can be observed that the graph subdivision stage dominates the overall computational cost.\   
Moreover,\   the refinement time is closely related to the number of connected components in the intersection point set,\   as well as the number of singular points.\   
Therefore,\   Examples~2,\ 8,\  and~9 exhibit the largest runtimes among all test examples.

\begin{table}[htbp]
    \centering 
    \caption{The intersection of two B-spline surfaces is divided into segments within the $(u, v)$ parameter domain. 
    OCCT refers to the intersection algorithm in the Open Cascade Technology. 
    Different segments are distinguished by different colors. }
    \label{tab:exp_compare}
    \begin{tabular}{m{1.4cm}<{\centering} | m{2.5cm}<{\centering} | m{2.5cm}<{\centering}}
        \toprule
        & OCCT & Ours  \\
        \midrule
        Example 3 &
        \includegraphics[width=0.15\textwidth]{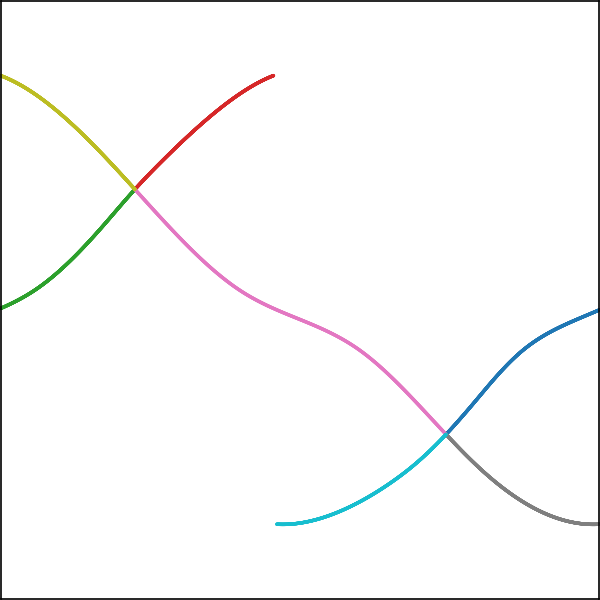 } &
        \includegraphics[width=0.15\textwidth]{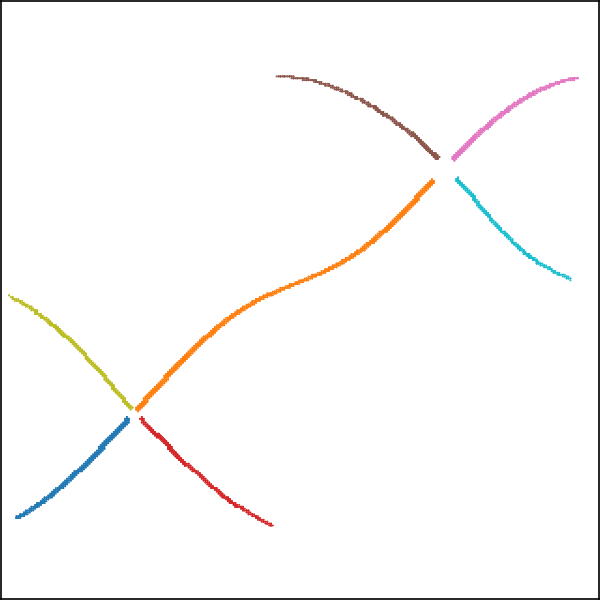}  
        \\ 
        Example 8 &
        \includegraphics[width=0.15\textwidth]{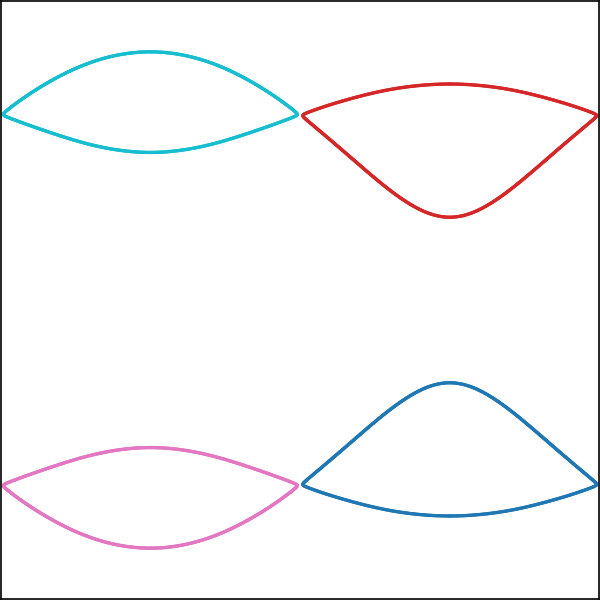 } &
        \includegraphics[width=0.15\textwidth]{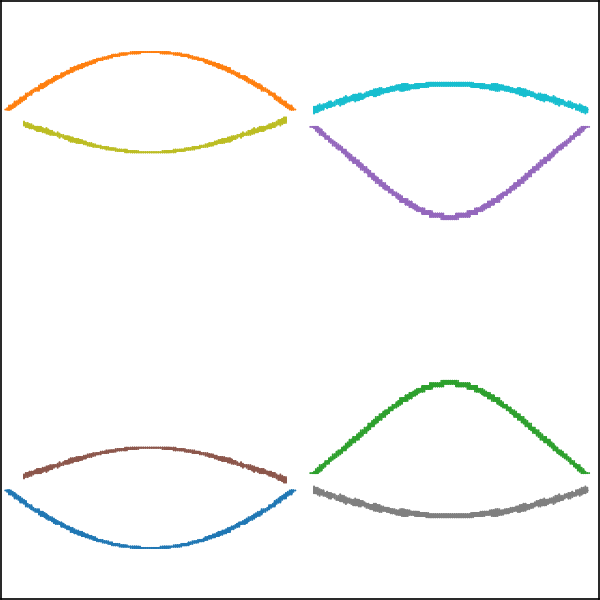} 
        \\ 
        Example 9 &
         \includegraphics[width=0.15\textwidth]{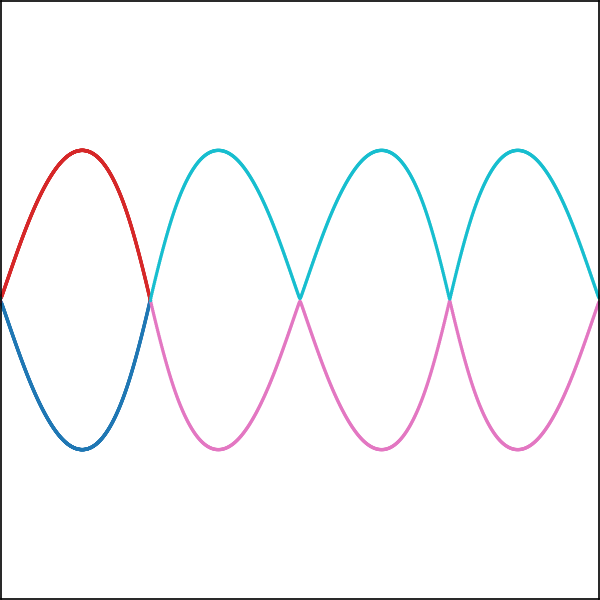} &
         \includegraphics[width=0.15\textwidth]{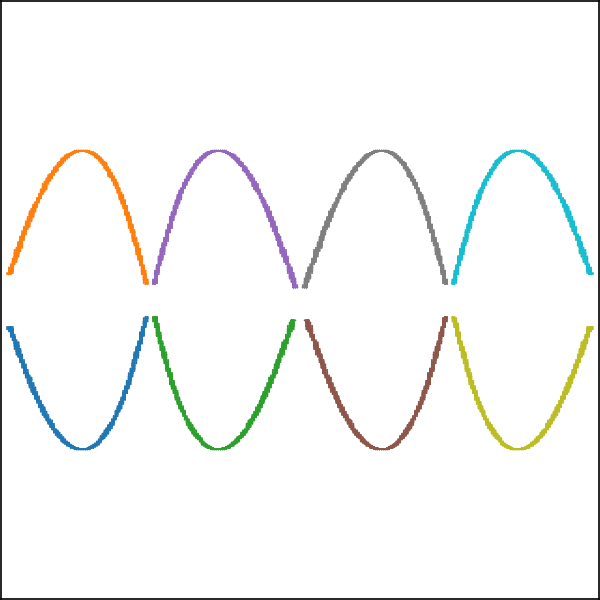} 
         \\
        \bottomrule
    \end{tabular}
\end{table}

\begin{table*}[htbp]
    \centering 
    \caption{Topology understanding of different surface/surface intersections$(\theta_{\text{ov}}=0.2)$.\   
    Corresponding results in different parameter domains are drawn for clarity.\   
    The first row shows the intersecting B-spline surfaces; 
    the second and third rows are the intersection point sets $P_{1,u,v}$ and $P_{2,s,t}$ ; 
    the fourth row is the Mapper graph of point set $P_{1,u,v}$; 
    the fifth and sixth rows are the results corresponding to $P_{1,u,v}$ and $P_{2,s,t}$. 
    Different intersection segments are represented in different colors. }
    \label{tab:exp_res1}
    \begin{tabular}{m{1.5cm}<{\centering}|m{2.5cm}<{\centering}|m{2.5cm}<{\centering}|m{2.5cm}<{\centering}|m{2.5cm}<{\centering}}
        \toprule
        & Example 1 & Example 2 & Example 3 & Example 4 \\
        \midrule
        Surfaces &
        \includegraphics[width=0.15\textwidth]{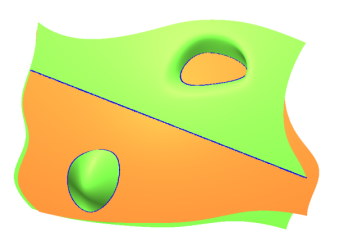} &
        \includegraphics[width=0.15\textwidth]{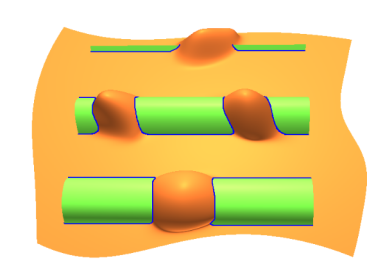} &
        \includegraphics[width=0.15\textwidth]{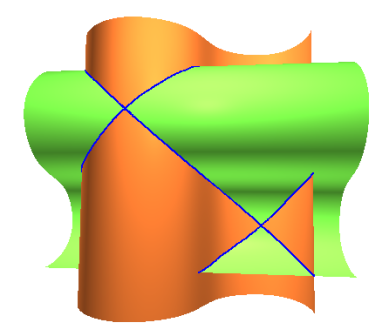} &
        \includegraphics[width=0.15\textwidth]{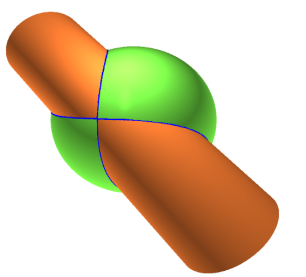} \\ 
        $P_{1,u,v}$ &
        \includegraphics[width=0.15\textwidth]{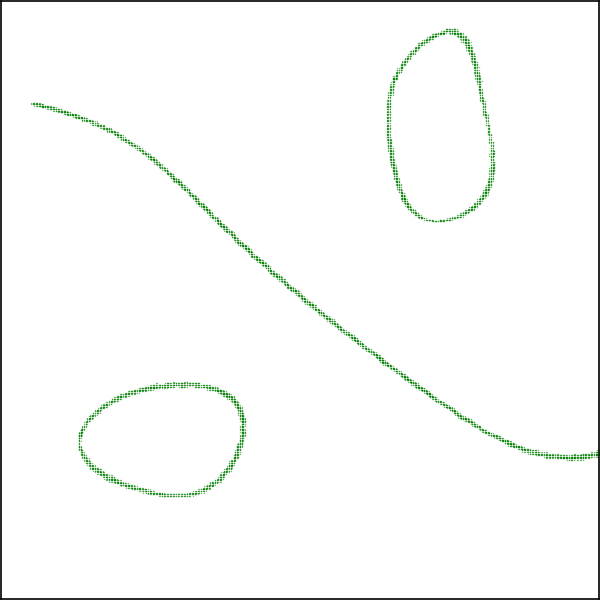} &
        \includegraphics[width=0.15\textwidth]{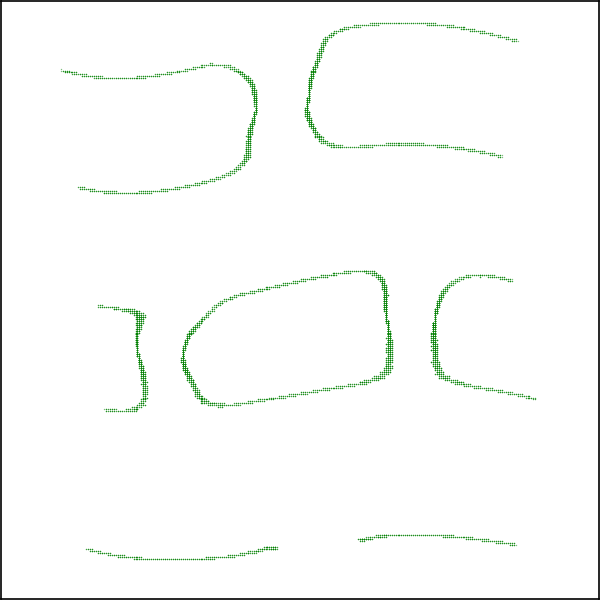} &
        \includegraphics[width=0.15\textwidth]{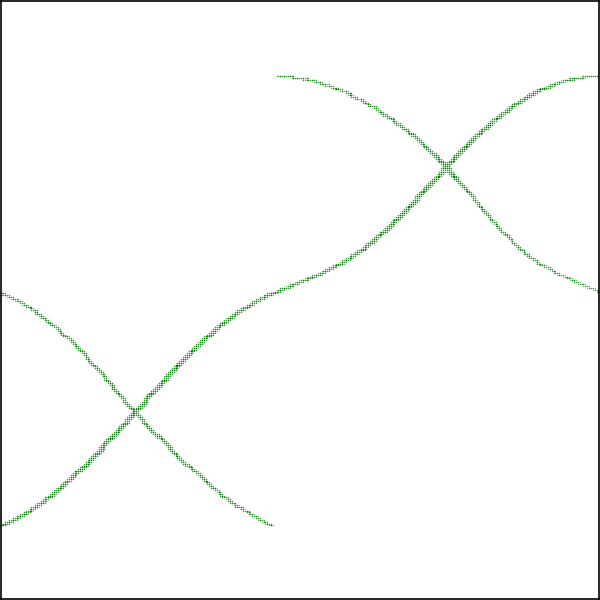} &
        \includegraphics[width=0.15\textwidth]{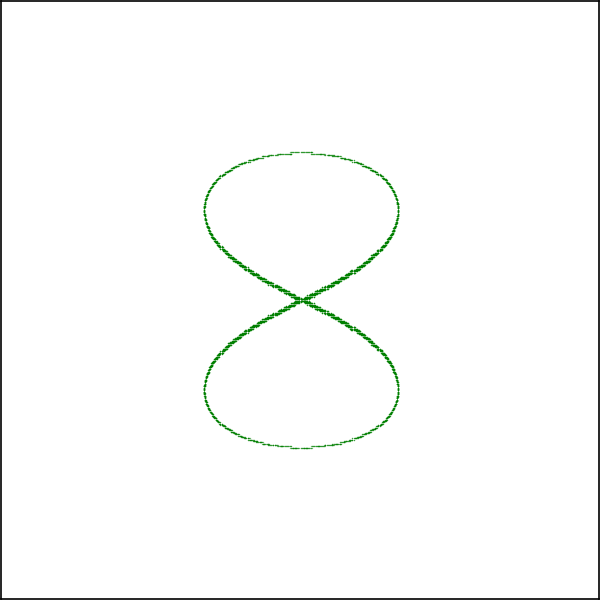} \\ 
        $P_{2,s,t}$ &
        \includegraphics[width=0.15\textwidth]{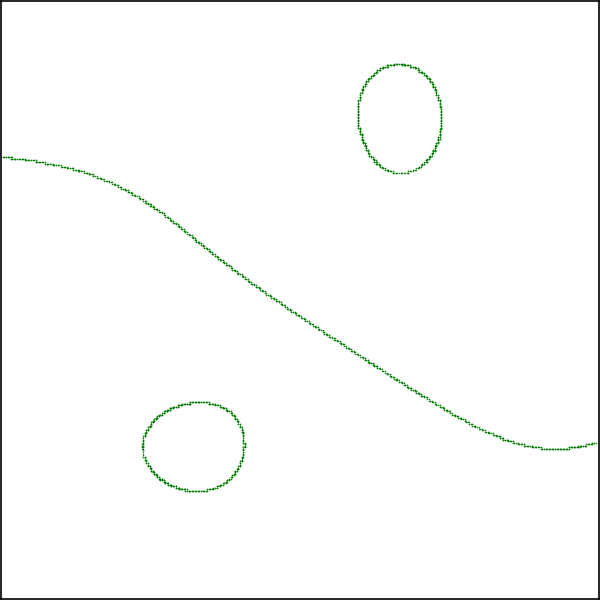} &
        \includegraphics[width=0.15\textwidth]{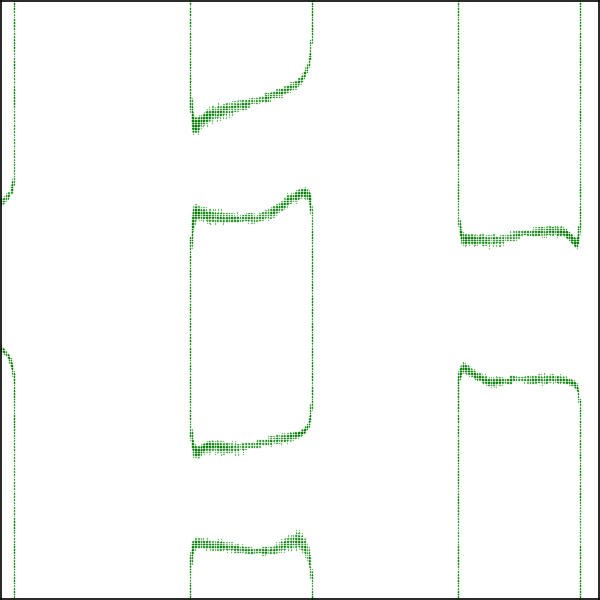} &
        \includegraphics[width=0.15\textwidth]{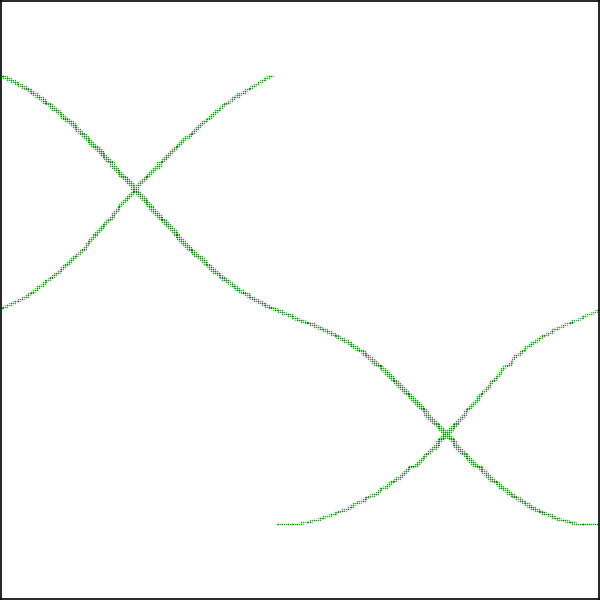} &
        \includegraphics[width=0.15\textwidth]{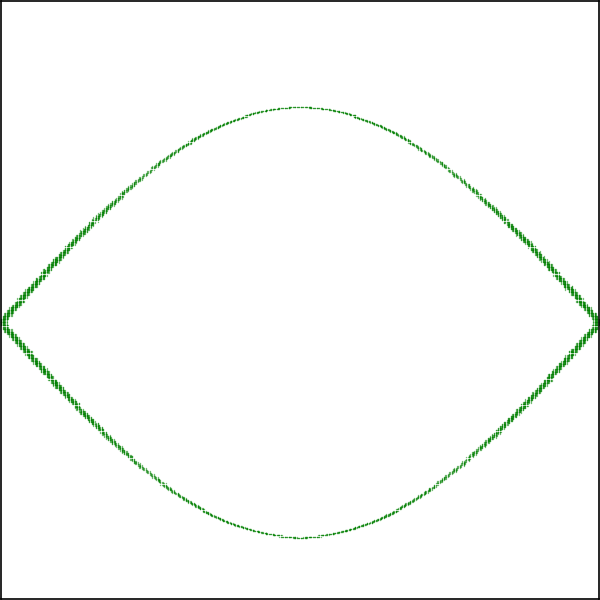} \\ 
        Mapper graph of $P_{1,u,v}$ &
        \includegraphics[width=0.15\textwidth]{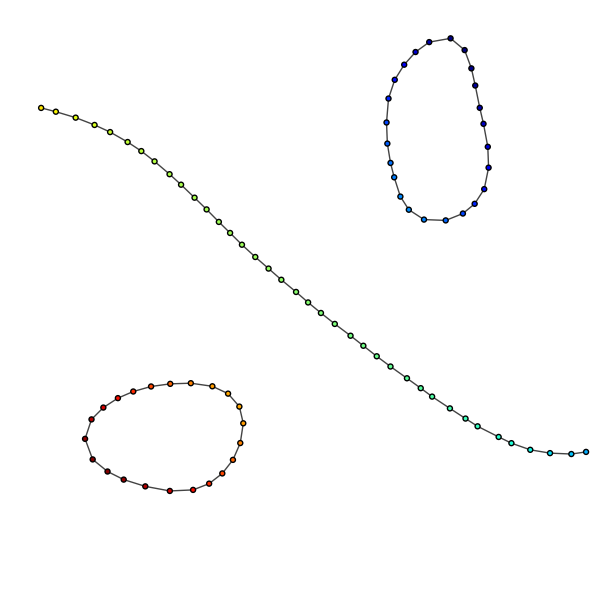} &
        \includegraphics[width=0.15\textwidth]{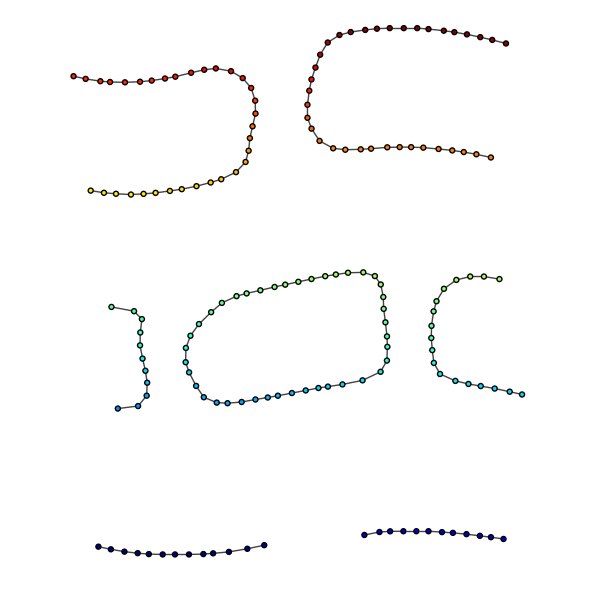} &
        \includegraphics[width=0.15\textwidth]{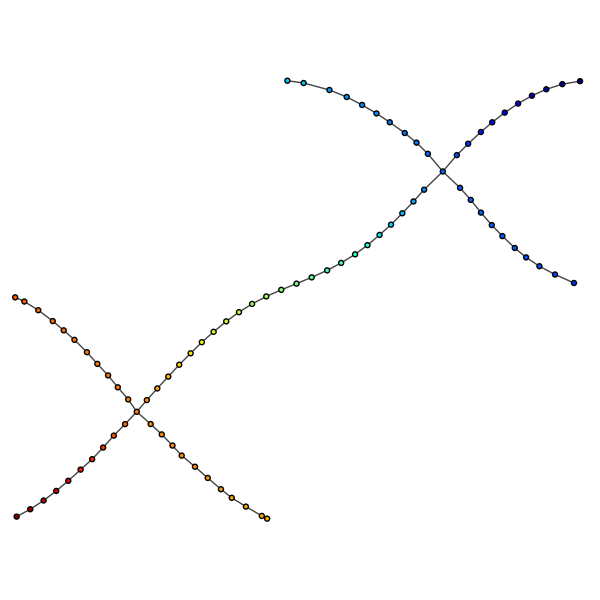} &
        \includegraphics[width=0.15\textwidth]{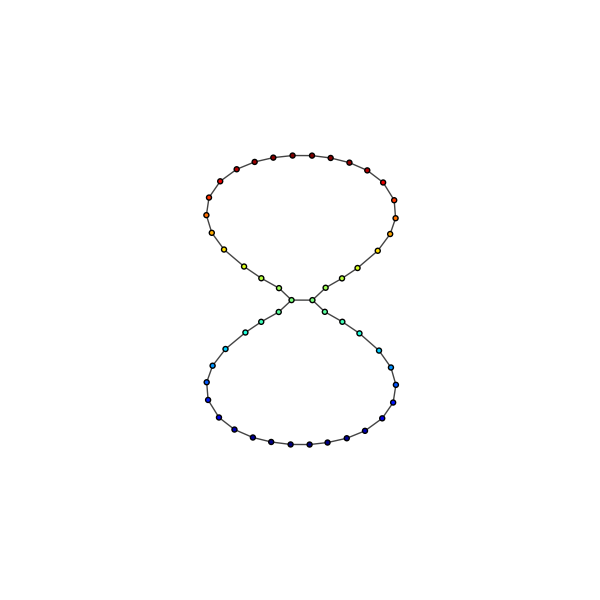} \\ 
        Result of $P_{1,u,v}$ &
        \includegraphics[width=0.15\textwidth]{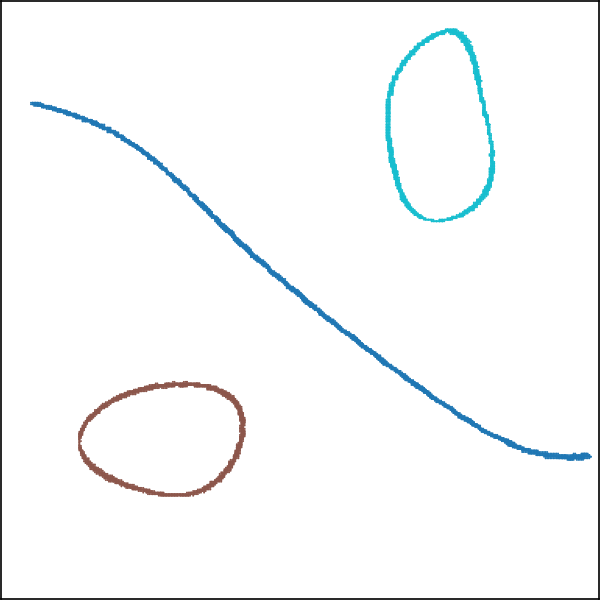} &
        \includegraphics[width=0.15\textwidth]{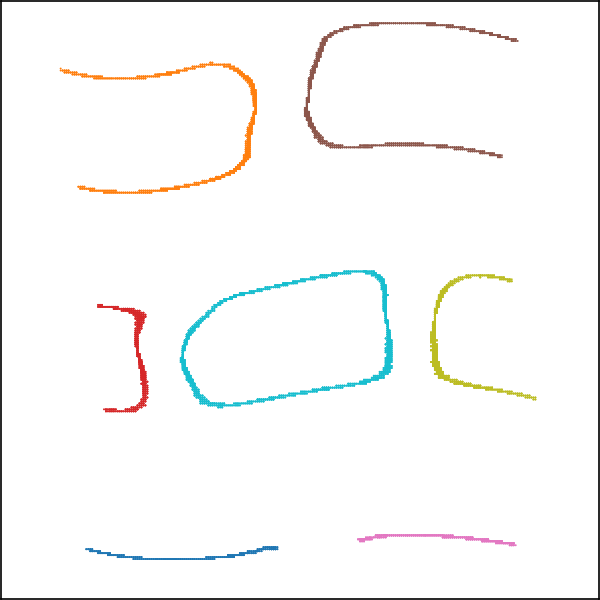} &
        \includegraphics[width=0.15\textwidth]{images/experiments/multi_jiaocha/multi_jiaocha_uv_res.png} &
        \includegraphics[width=0.15\textwidth]{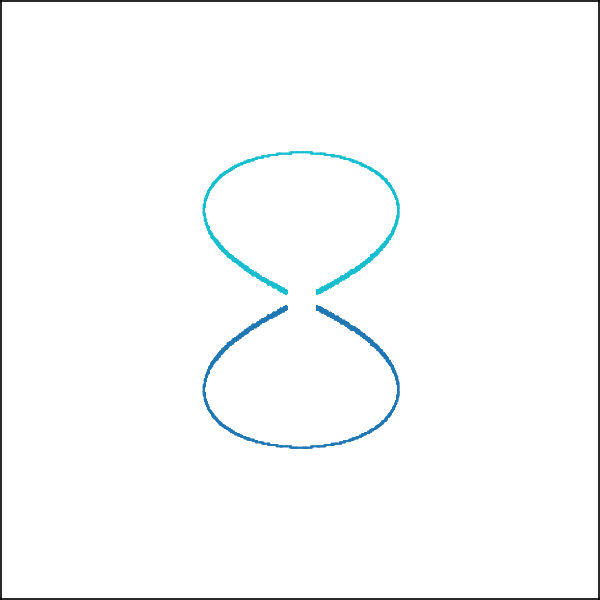} \\ 
        Result of $P_{2,s,t}$ &
        \includegraphics[width=0.15\textwidth]{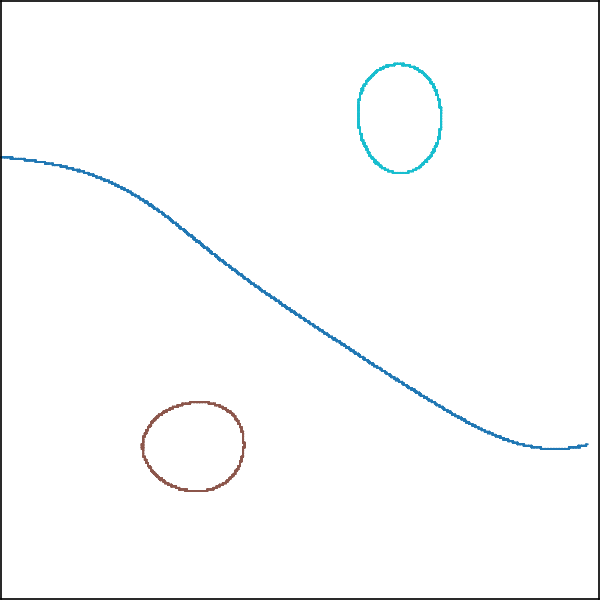} &
        \includegraphics[width=0.15\textwidth]{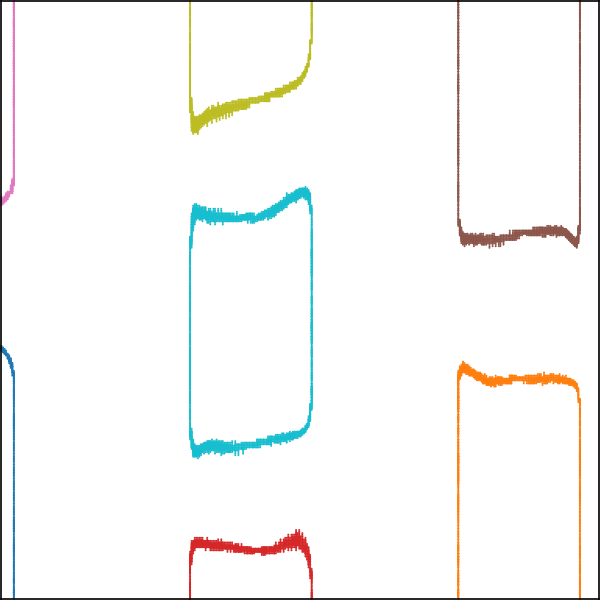} &
        \includegraphics[width=0.15\textwidth]{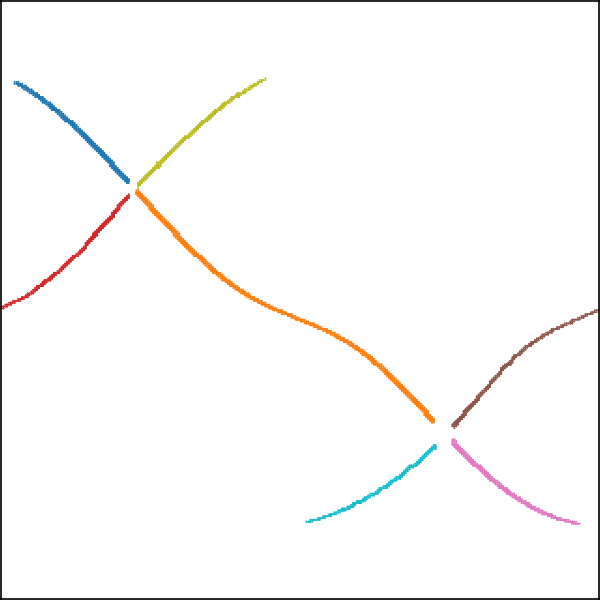} &
        \includegraphics[width=0.15\textwidth]{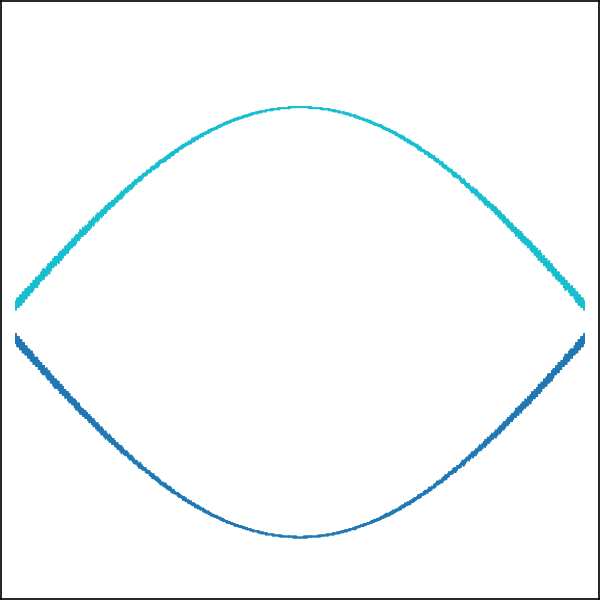} \\ 
        \bottomrule
    \end{tabular}
\end{table*}

\begin{table*}[htbp]
    \centering 
    \caption{More results of different surface/surface intersections$(\theta_{\text{ov}}=0.2)$.\   The arrangement of content in each row and column remains consistent with Table \ref{tab:exp_res1}}
    \label{tab:exp_res2}
    \begin{tabular}{m{1.5cm}<{\centering}|m{2.5cm}<{\centering}|m{2.5cm}<{\centering}|m{2.5cm}<{\centering}|m{2.5cm}<{\centering}|m{2.5cm}<{\centering}}
        \toprule
        & example 5 &Example 6 & Example 7 & Example 8 & Example 9 \\
        \midrule
        Surfaces &
        \includegraphics[width=0.15\textwidth]{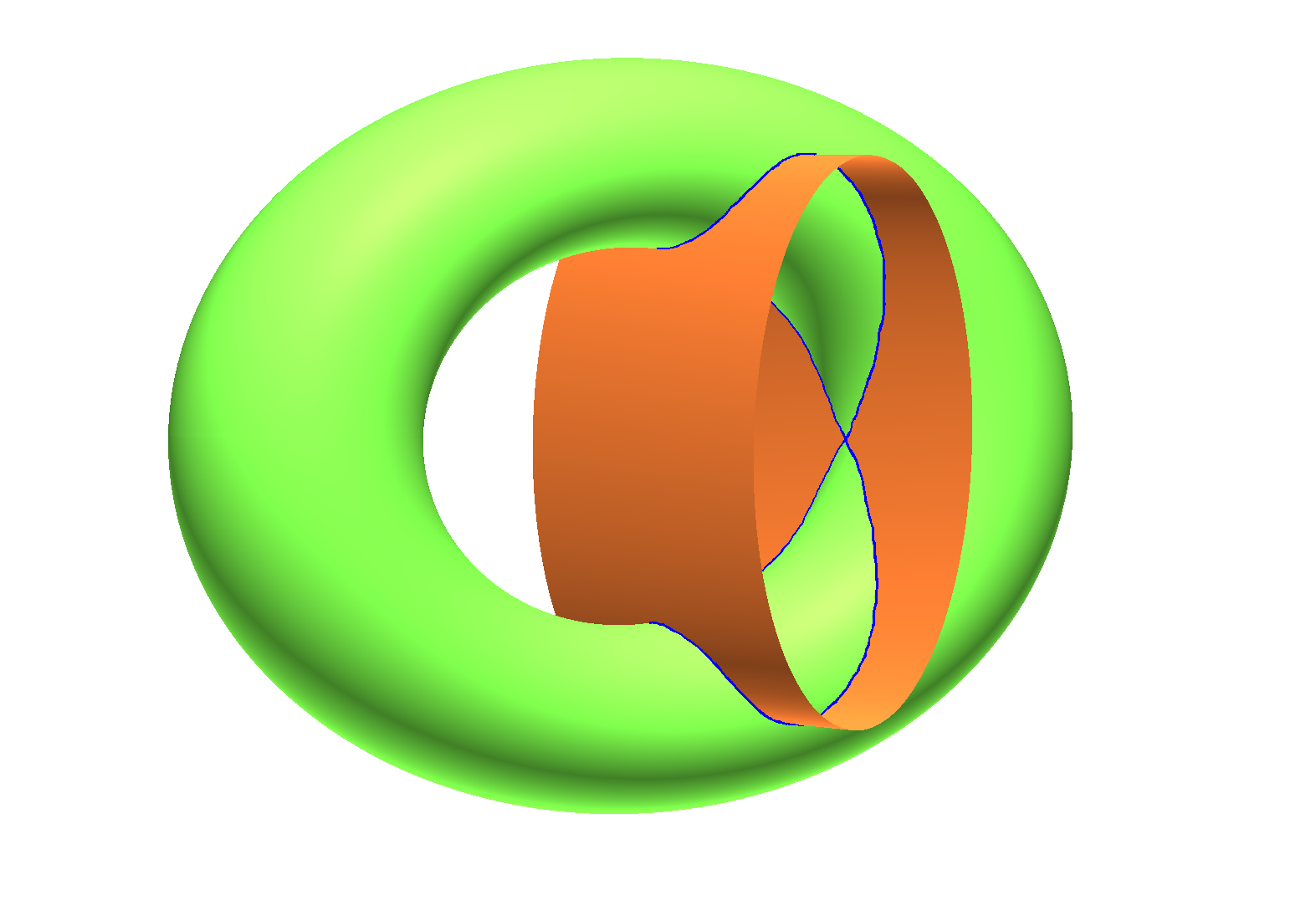} &
        \includegraphics[width=0.15\textwidth]{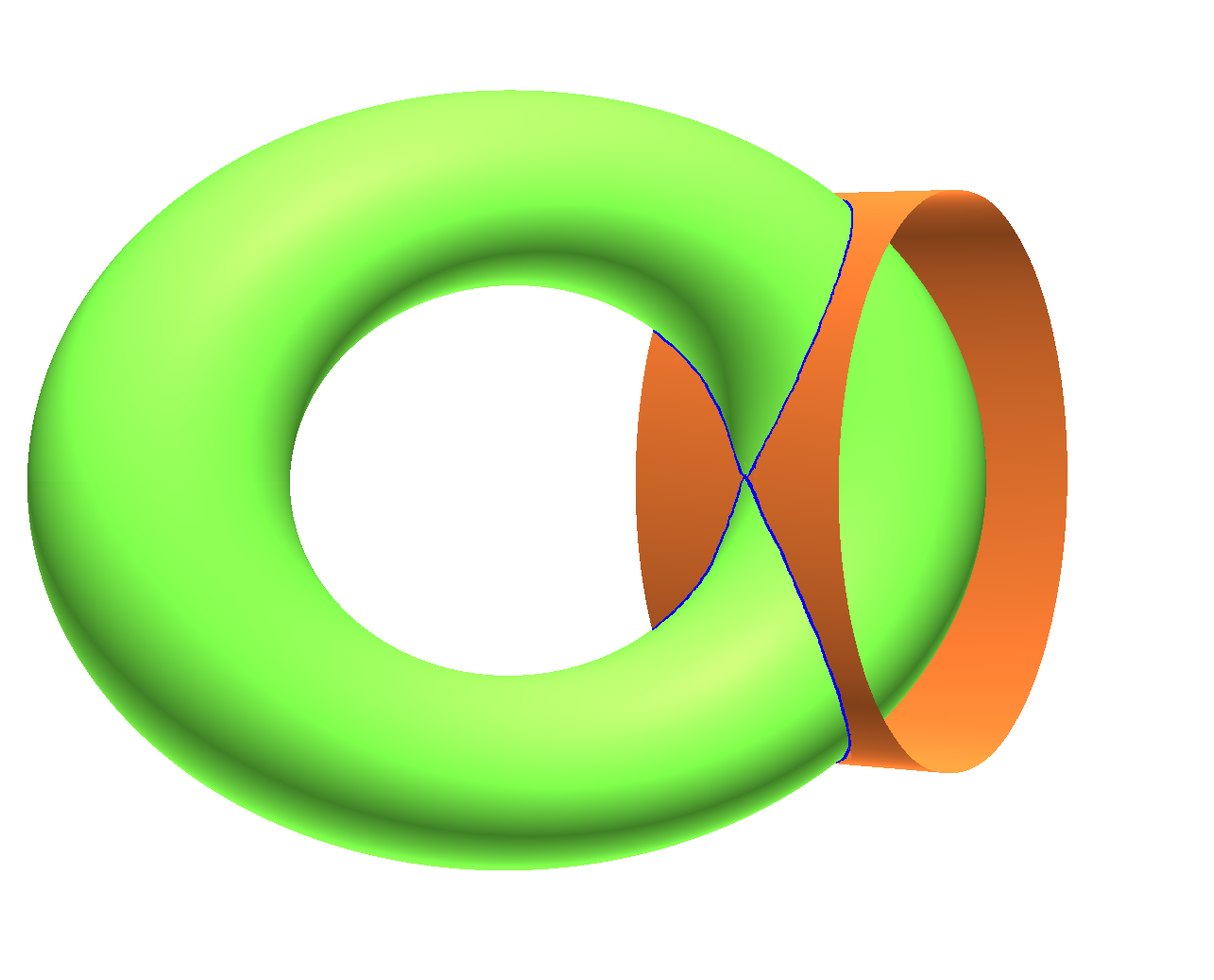} &
        \includegraphics[width=0.15\textwidth]{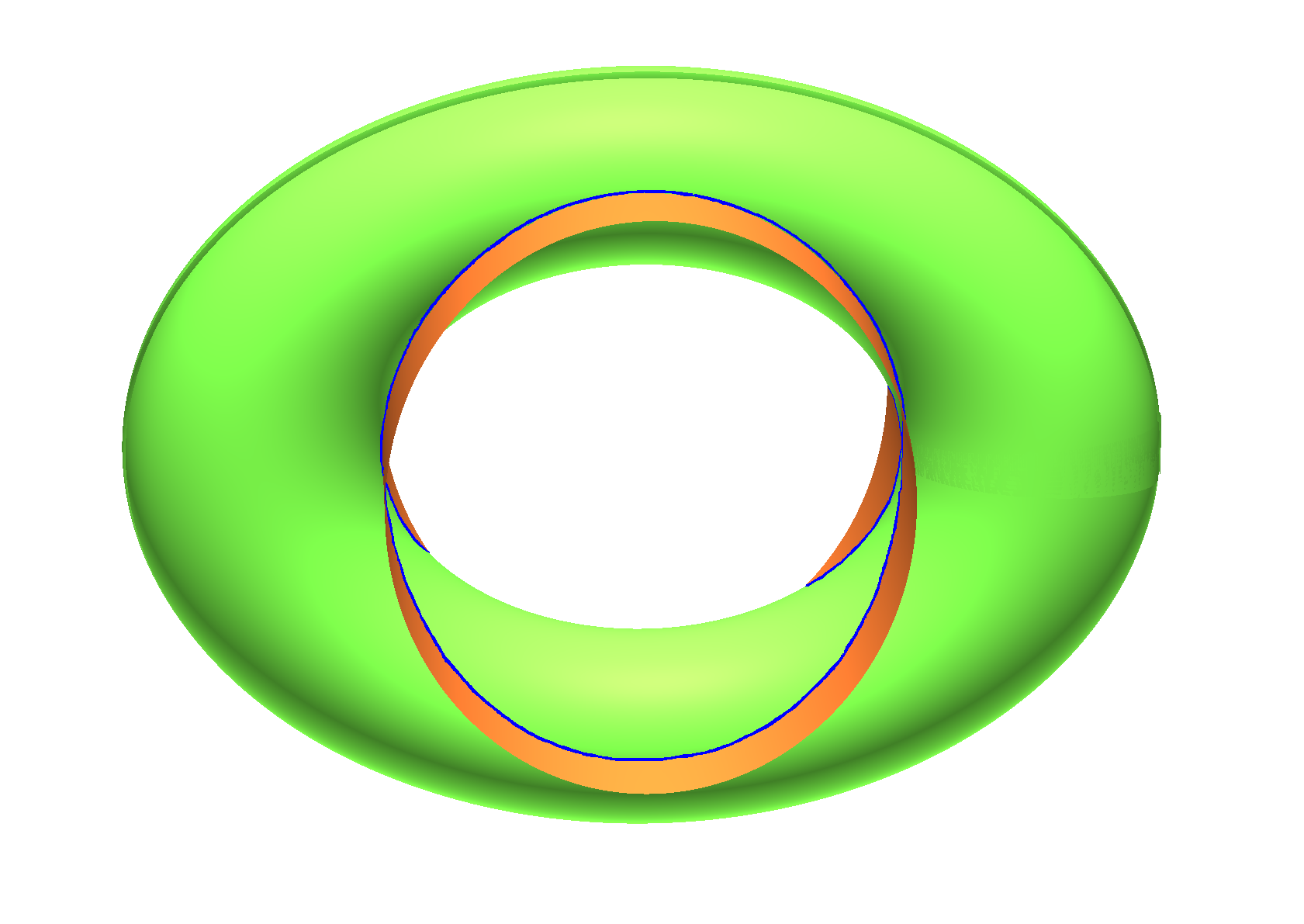} &
        \includegraphics[width=0.15\textwidth]{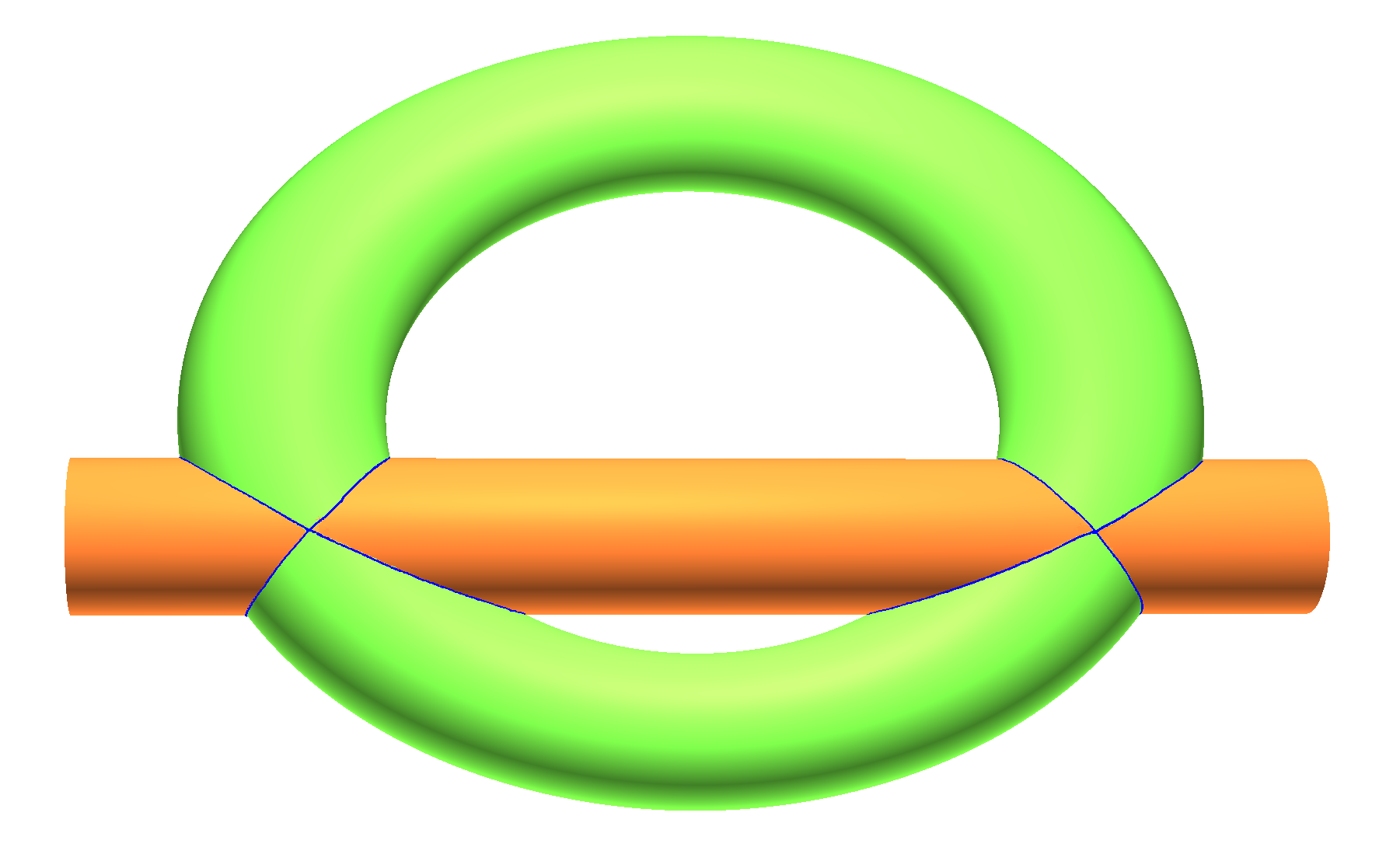} &
        \includegraphics[width=0.15\textwidth]{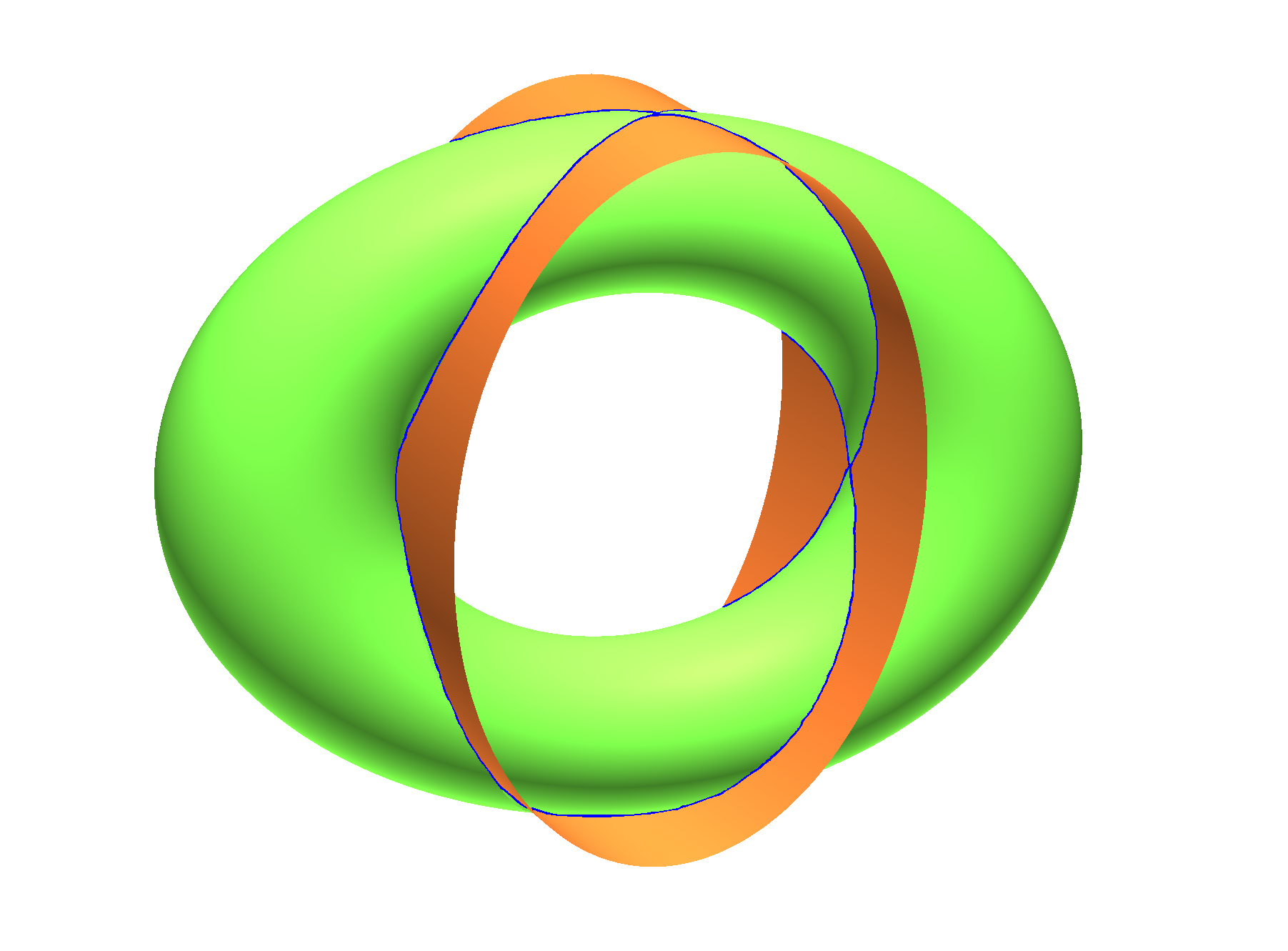} \\ 
        $P_{1,u,v}$ &
        \includegraphics[width=0.15\textwidth]{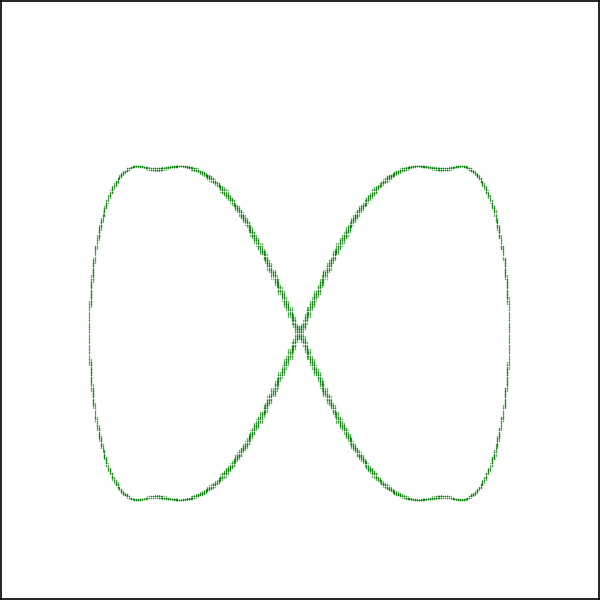} &
        \includegraphics[width=0.15\textwidth]{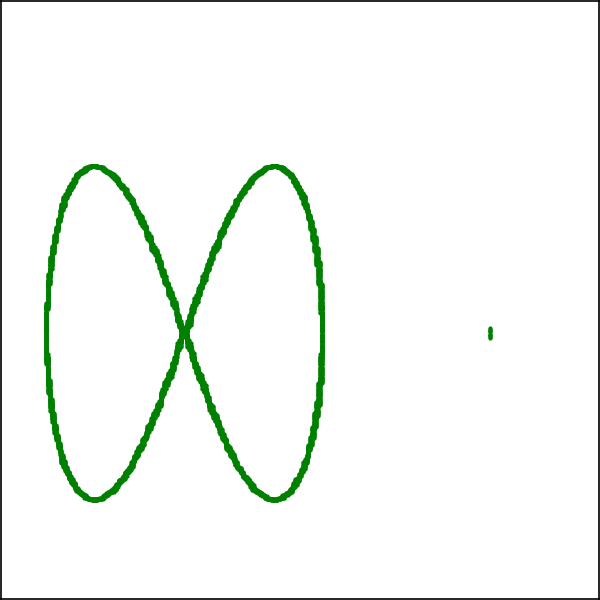} &
        \includegraphics[width=0.15\textwidth]{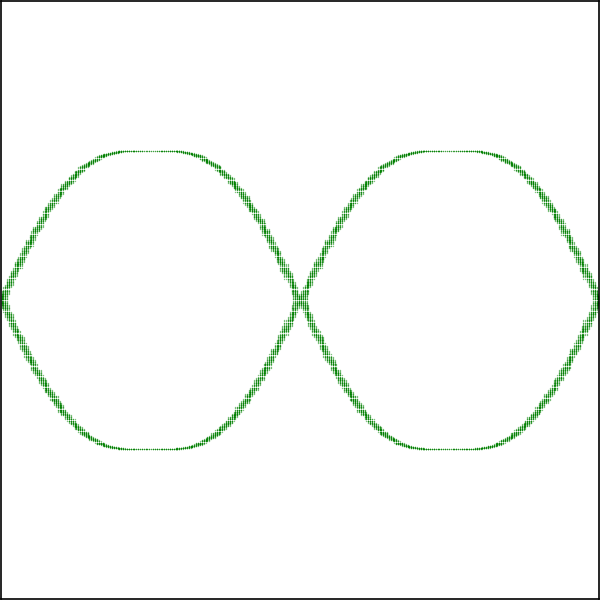} &
        \includegraphics[width=0.15\textwidth]{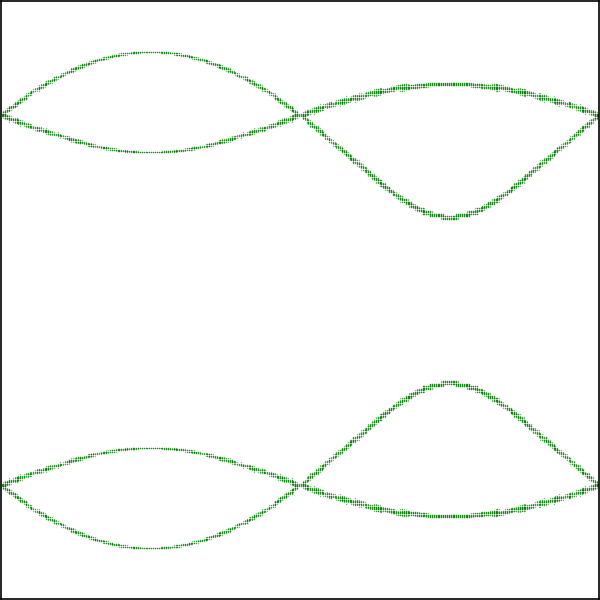} &
        \includegraphics[width=0.15\textwidth]{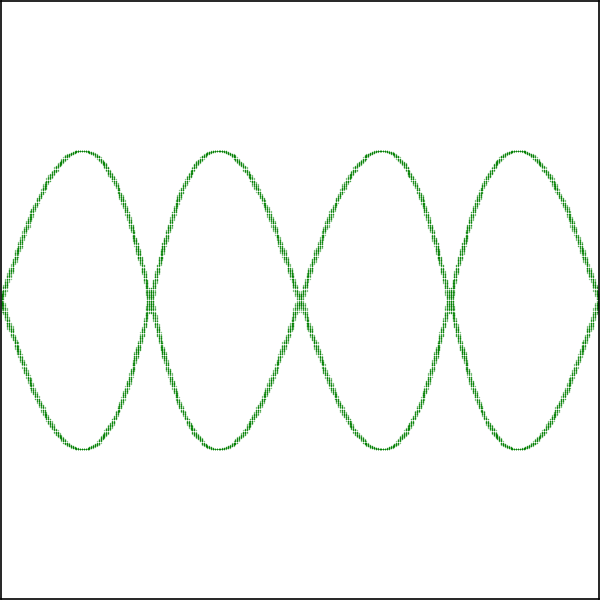} \\ 
        $P_{2,s,t}$ &
        \includegraphics[width=0.15\textwidth]{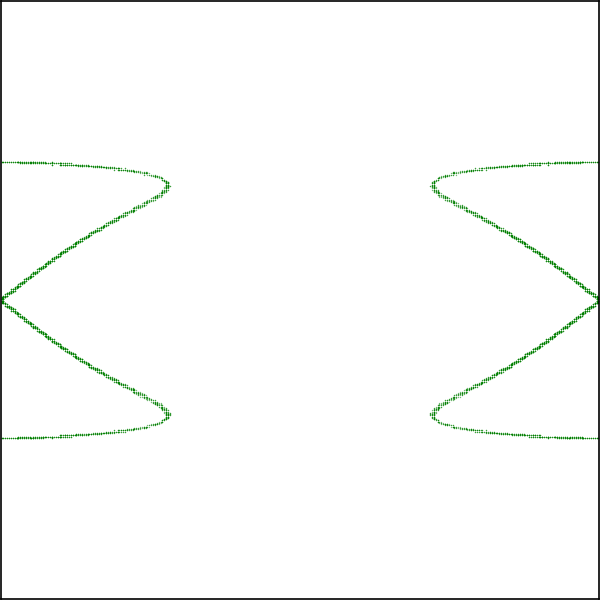} &
        \includegraphics[width=0.15\textwidth]{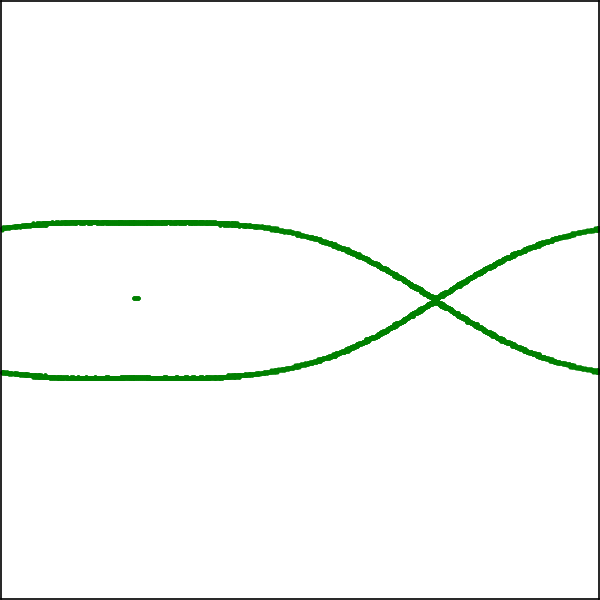} &
        \includegraphics[width=0.15\textwidth]{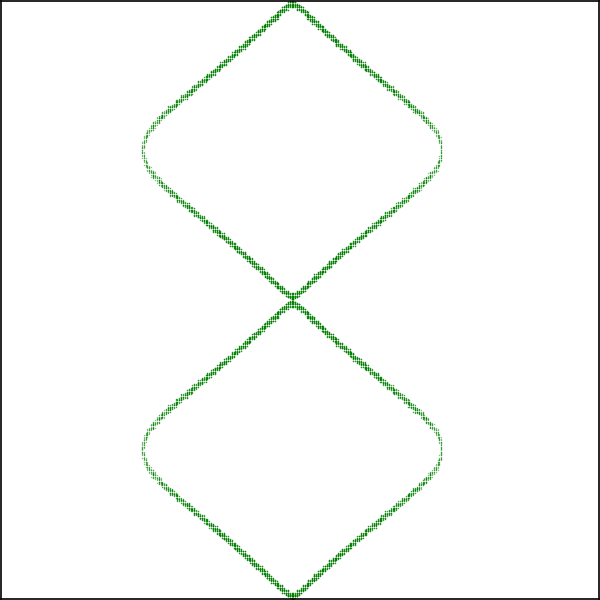} &
        \includegraphics[width=0.15\textwidth]{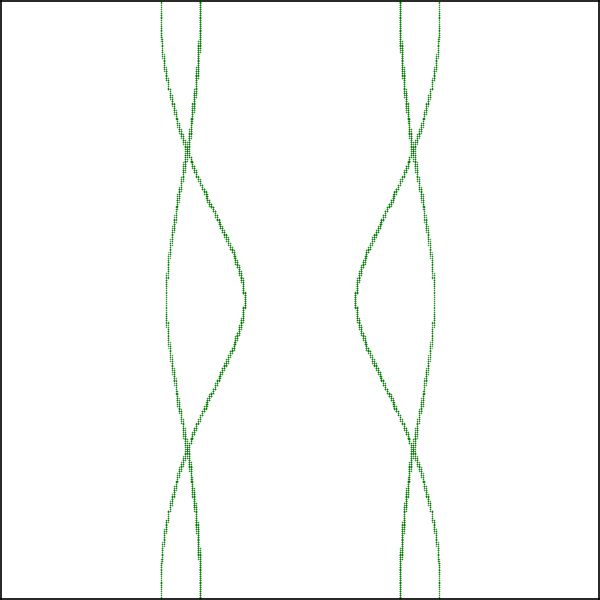} &
        \includegraphics[width=0.15\textwidth]{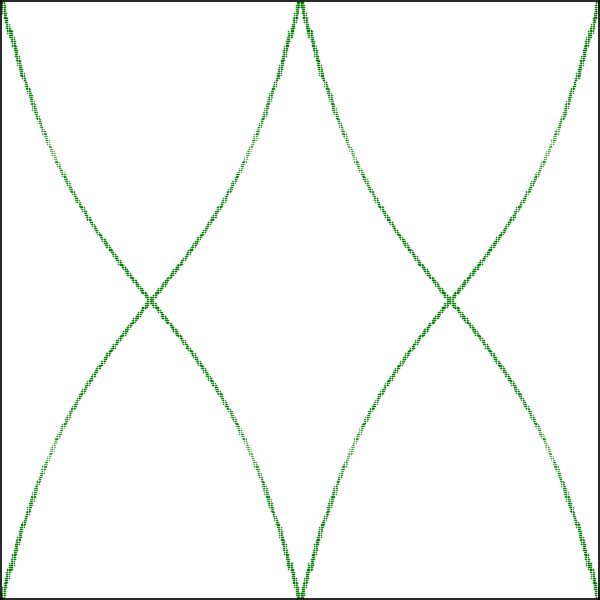} \\ 
        Mapper graph of $P_{1,u,v}$ &
        \includegraphics[width=0.15\textwidth]{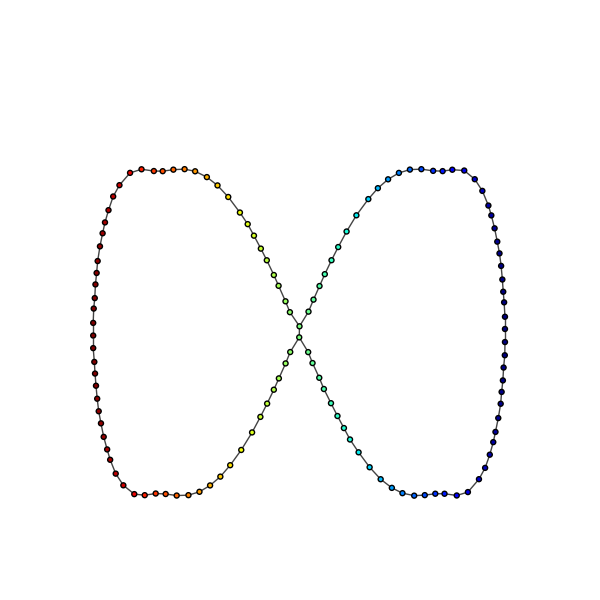} &
        \includegraphics[width=0.15\textwidth]{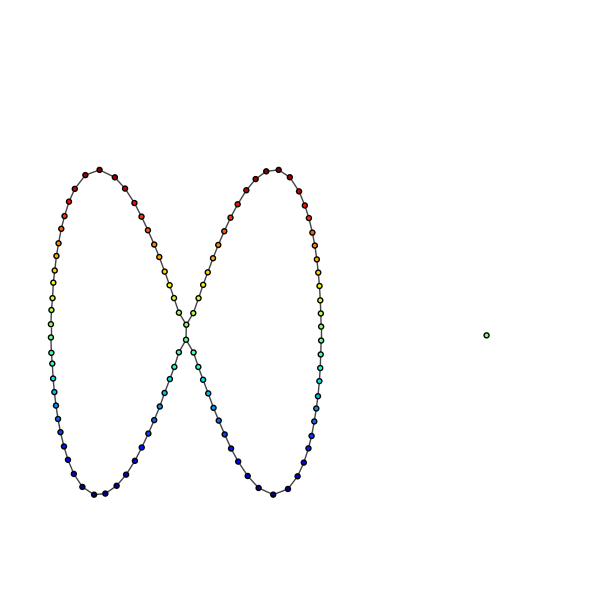} &
        \includegraphics[width=0.15\textwidth]{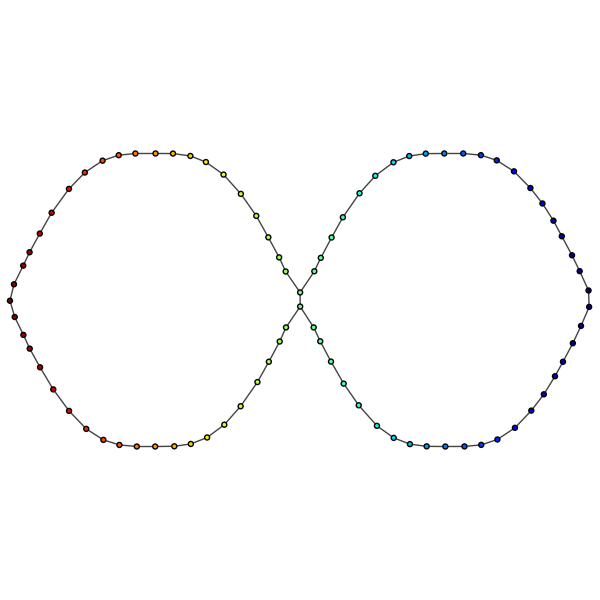} &
        \includegraphics[width=0.15\textwidth]{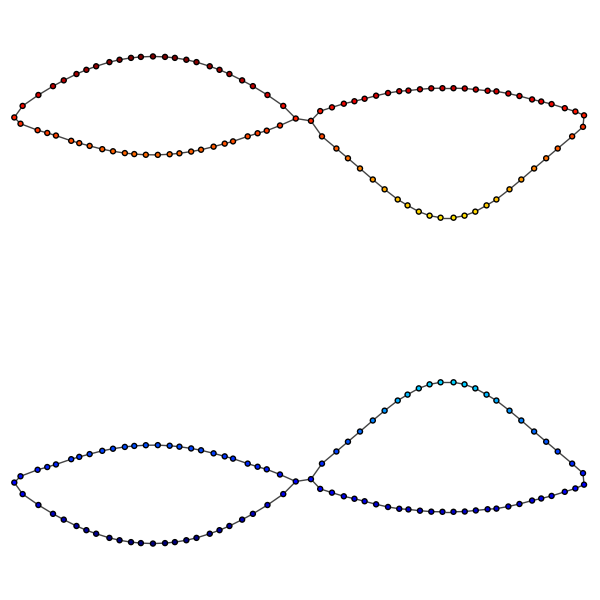} &
        \includegraphics[width=0.15\textwidth]{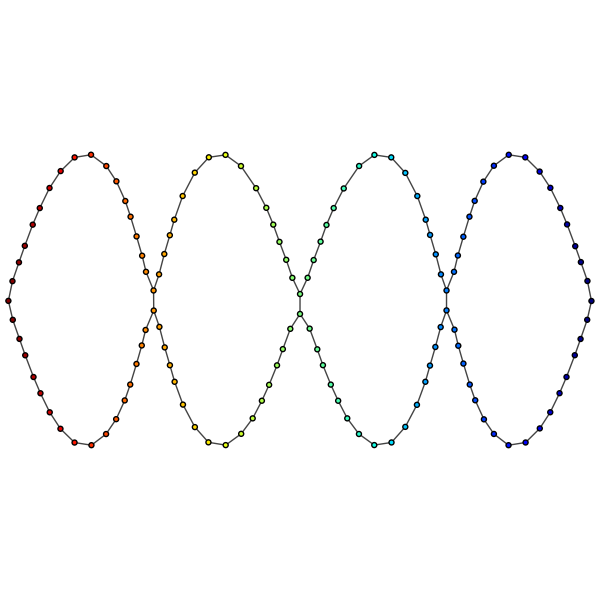} \\ 
        Result of $P_{1,u,v}$ &
        \includegraphics[width=0.15\textwidth]{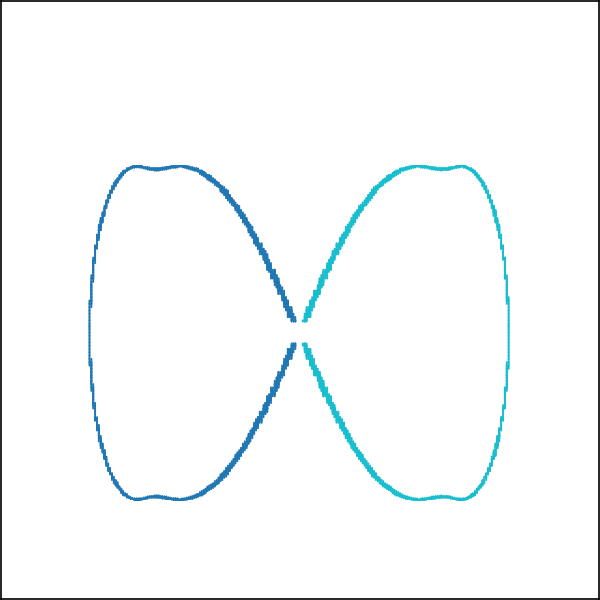} &
        \includegraphics[width=0.15\textwidth]{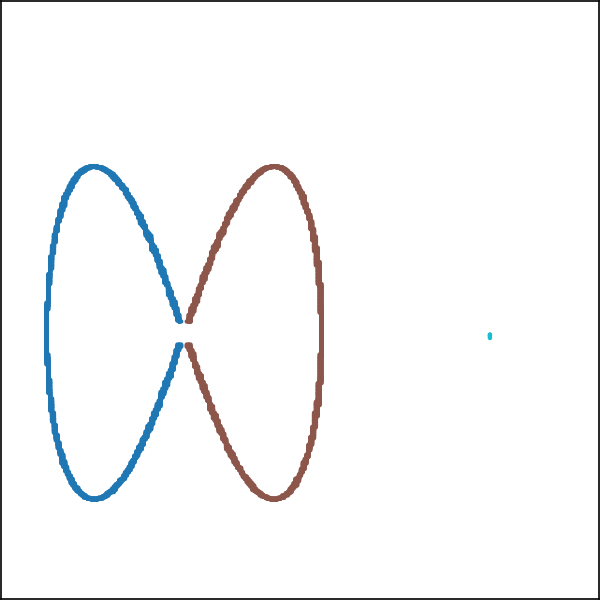} &
        \includegraphics[width=0.15\textwidth]{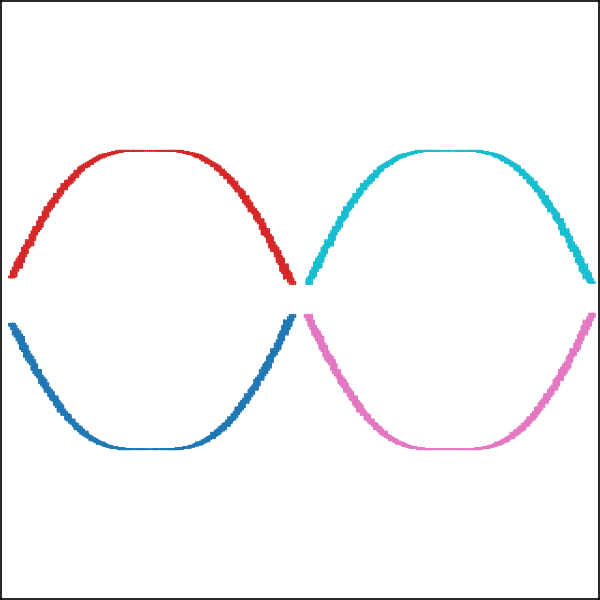} &
        \includegraphics[width=0.15\textwidth]{images/experiments/torus_cylinder1/torus_cylinder1_uv_res.png} &
        \includegraphics[width=0.15\textwidth]{images/experiments/torus_cylinder6/torus_cylinder6_uv_res.png} \\ 
        Result of $P_{2,s,t}$ &
        \includegraphics[width=0.15\textwidth]{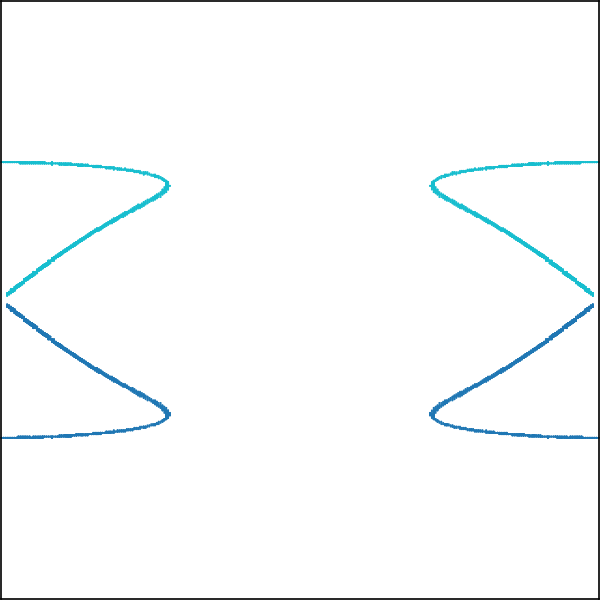} &
        \includegraphics[width=0.15\textwidth]{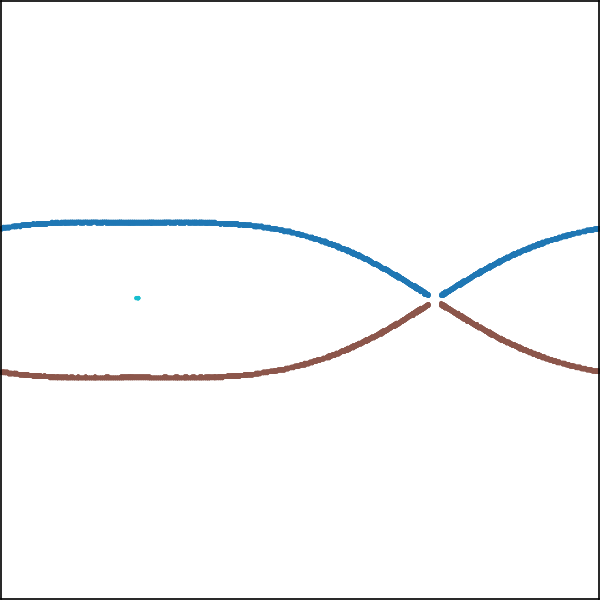} &
        \includegraphics[width=0.15\textwidth]{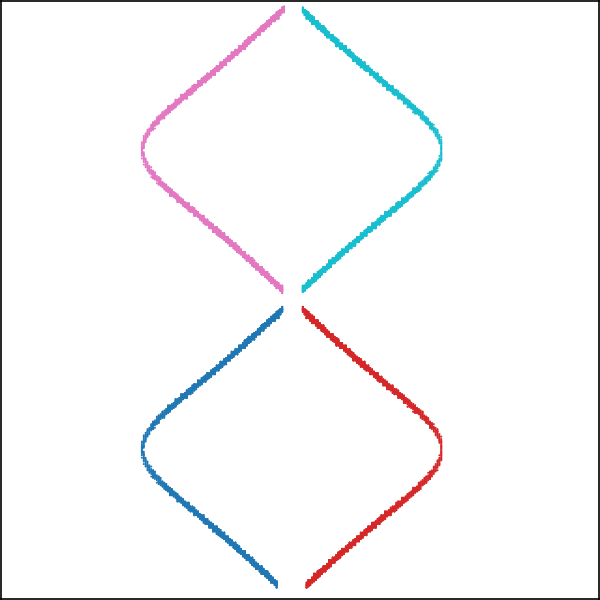} &
        \includegraphics[width=0.15\textwidth]{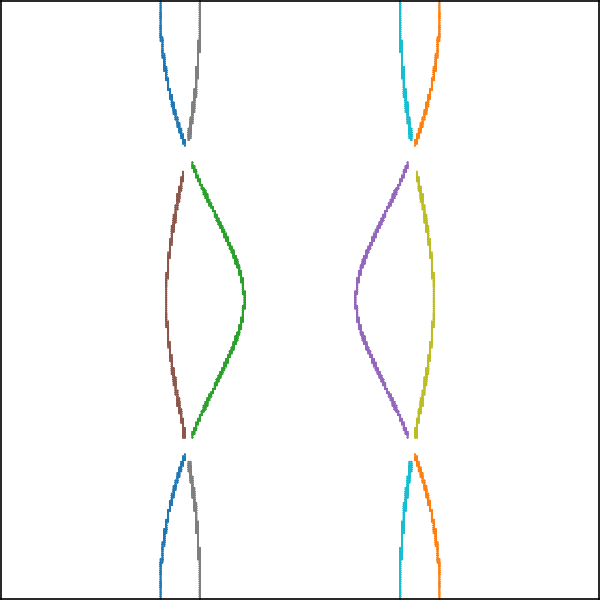} &
        \includegraphics[width=0.15\textwidth]{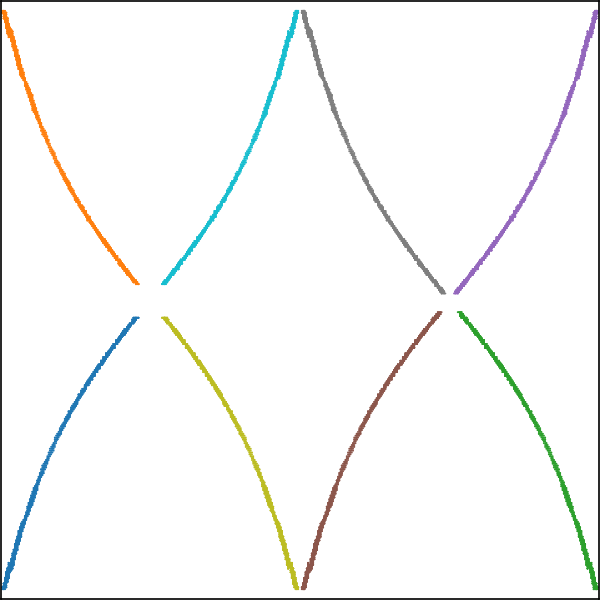} \\ 
        \bottomrule
    \end{tabular}
\end{table*}

\textbf{Multiple connected branches}.\   We first consider the case where two intersecting surfaces contain multiple connected components.\   
In Example 1 and Example 2 in Table \ref{tab:exp_res1},\  the intersection of these two surfaces contain multiple  connected components.\   The results in Table \ref{tab:exp_res1} show that our algorithm  can correctly identify and distinguish all connected branches of the intersection,\   and obtain a one-to-one correspondence between the parametric domains of the two surfaces.\

\textbf{Singular points}.\   We also investigated the performance of the algorithm when dealing with  intersections containing singular points.\    In the example 3 in Table \ref{tab:exp_res1} ,\   the intersection of two B-spline surfaces contains two singular points.\    Our algorithm can detect the location of the two singular points and split the intersection point sets into some simple subsets at the singular points and keep the topology consistent across different parameter domains.\

\textbf{Different topology in different parameter domains}.\   Even when the topology of the intersection of different parameter domains is different,\   our method can still obtain the correct correspondence.\    Example 4 in Table \ref{tab:exp_res1} and Example 5 in Table \ref{tab:exp_res2} show the cases that two periodic B-spline surfaces intersecting.\  Due to periodicity,\    we need to partition the intersection set into different subsets at the boundary points in the parameter domain.\     Since the two surfaces $B_1(u,\  v)$ and $B_2(s,\  t)$ are unfolded at different positions in the parameter domain,\   the topology of the intersection point sets $P_{1,u,v}$ and $P_{2,s,t}$ of the parameter domain is different.\    However,\   we can still obtain the correct correspondence.\   

\textbf{Intersection with isolated points}.\     Our method can also precisely identify isolated intersection points.\     Example 6 in Table \ref{tab:exp_res2} demonstrates a case where two surfaces intersect at isolated tangent points.\     By locating nodes with degree 0 in the Mapper graph, our method can distinguish isolated tangent contact points and establish the correct correspondence between the parameter domains of the two surfaces.\    

\textbf{Combinations of multiple situations}.\     Even when intersections involve combinations of the aforementioned complex cases, our method achieves reliable topological understanding.\    In Example 7 in Table \ref{tab:exp_res2}, the intersection of the two surfaces contains both singular points and periodic boundary points.\     In Example 8 in Table \ref{tab:exp_res2}, the intersection of the two surfaces contains singular points and periodic boundary points, and  multiple connected components.\     In Example 9 in Table \ref{tab:exp_res2}, the intersection of the two surfaces contains multiple singular points and periodic boundary points.\     Our method can effectively identifies their topological structures,\    correctly partition intersection point sets, and establishes accurate correspondences between the parameter domains of two intersecting surfaces.\      This validates the effectiveness of our method in handling complex topology.\    

\subsection{Comparison}

To evaluate the effectiveness of the proposed method, we compare it with the intersection-processing module in the open-source software Open CASCADE Technology (OCCT). 
The version of OCCT employed in this work is 7.8.1. 
Table \ref{tab:exp_compare} summarizes the results for examples containing multiple singular points, with the ordering consistent with Tables \ref{tab:exp_res2} and \ref{tab:exp_res1}.

In example 3, both methods produce seven segments. 
In Example 8, OCCT generates only four segments because surface periodicity is not considered, whereas the proposed method produces eight segments. 
In Example 9, OCCT partitions the intersection into four segments, resulting in tangential discontinuities. 
In contrast, the proposed method separates the curve at all singular points and produces eight geometrically continuous segments.

Overall, compared with OCCT, the proposed method achieves a finer and topologically consistent segmentation of the intersection. 
Although OCCT can partition the intersection into non-self-intersecting segments, it does not explicitly address topological transitions or geometric continuity at singular points. 
As a result, the intersection may not be fully separated at singularities, leading to discontinuities in tangent direction.
By incorporating global topological information, the proposed method produces a finer partition in which each segment is free of self-intersection and maintains tangential continuity.

\subsection{Discussion and limitations}

 Despite the promising results,\   the proposed algorithm has several limitations.\    One of the main limitations is related to the handling of clustered singular points.\    When multiple singular points are in close proximity,\   the current approach of constructing interval coverages for the filter function in the Mapper graph may assign these points to the same interval.\    As a result,\   they are clustered together as a single node in the Mapper graph,\   causing the algorithm to incorrectly identify them as a single singular point.\    
\section{Conclusion}\label{sec:Conclusion}

 In this paper,\  we present a new algorithm for understanding the topology of intersections between B-spline surfaces with Mapper.\    
The algorithm first constructs a Mapper graph using a Two-step Mapper algorithm.\     
Next,\   the algorithm identifies characteristic nodes in the Mapper graph.\   
Finally,\   by partitioning the Mapper graph at these characteristic nodes,\   we divide the intersections into simpler segments,\   enabling a more desirable understanding  of the topological structure of the intersections.\      

The significance of our method lies in its ability to determine the overall topology of the intersection, which is fundamental to subsequent processes of surface/surface intersection calculation. 
The accurate topology understanding provided by our algorithm can enhance the robustness and efficiency of workflows in surface/surface intersection.\   
\section*{Declaration of competing interest}

The authors declare that they have no known competing financial interests or personal relationships that could have appeared
to influence the work reported in this paper.\   


\clearpage 

\bibliographystyle{unsrtnat}

\bibliography{cas-refs}

\end{document}